\documentclass[aps,prb,twocolumn,superscriptaddress]{revtex4-2}  

\usepackage{amsmath,amssymb,amsfonts,bm,physics,latexsym,comment} 

\usepackage[pdftex]{graphicx,color}
\usepackage{hyperref}
\hypersetup{
   colorlinks=true,        
   linkcolor=blue,          
   citecolor=blue,        
  filecolor=magenta,      
   urlcolor=blue           
}
\usepackage{tikz}

\newcommand{\tred}[1]{\textcolor{red}{#1}}

\newcommand{\Jperp}{J_\perp}
\newcommand{\Jdiag}{J_\times}
\newcommand{\Jpd}{J_{\perp,\times}}
\newcommand{\gt}{\tilde{g}}

\newcommand{\Sv}{\bm{S}}
\newcommand{\Mv}{\bm{M}}
\newcommand{\Nv}{\bm{N}}
\newcommand{\qv}{\bm{q}}

\newcommand{\Ocal}{{\cal O}}
\newcommand{\bs}{\mathrm{bs}}

\newcommand{\db}{\bar{d}}
\newcommand{\sqtp}{\sqrt{2\pi}}

\newcommand{\PBC}{\mathrm{PBC}}
\newcommand{\TBC}{\mathrm{TBC}}
\newcommand{\even}{\mathrm{even}}
\newcommand{\odd}{\mathrm{odd}}
\newcommand{\Erm}[1]{E_\mathrm{#1}}

\begin{document}
\title{
Ground-state phase diagram of a spin-$\frac12$ frustrated XXZ ladder
}
\author{Takuhiro Ogino}
\affiliation{Institute for Solid State Physics, University of Tokyo, Kashiwa, Chiba 277-8581, Japan}
\author{Ryui Kaneko}
\affiliation{Department of Physics, Kindai University, Higashi-Osaka, Osaka 577-8502, Japan}
\author{Satoshi Morita}
\affiliation{Institute for Solid State Physics, University of Tokyo, Kashiwa, Chiba 277-8581, Japan}
\author{Shunsuke Furukawa}
\affiliation{Department of Physics, Keio University, Kohoku-ku, Yokohama, Kanagawa 223-8522, Japan}

\date{\today}

\begin{abstract}
We study the ground-state phase diagram of a spin-$\frac12$ frustrated XXZ ladder, 
in which two antiferromagnetic chains are coupled by competing rung and diagonal interactions, $\Jperp$ and $\Jdiag$. 
Previous studies on the isotropic model have revealed that 
a fluctuation-induced effective dimer attraction between the legs stabilizes the columnar dimer (CD) phase 
in the highly frustrated regime $\Jperp\approx 2\Jdiag$, 
especially for ferromagnetic $\Jpd<0$. 
By means of effective field theory and numerical analyses, 
we extend this analysis to the XXZ model, and obtain a rich phase diagram. 
The diagram includes four gapped featureless phases with no symmetry breaking: 
the rung singlet (RS) and Haldane phases as well as their twisted variants, the RS* and Haldane* phases, 
which are all distinct in the presence of certain symmetries. 
Significantly, the Haldane-CD transition point in the isotropic model turns out to be a crossing point of two transition lines in the XXZ model, 
and the stripe N\'eel and RS* phases appear between these lines. 
This indicates a nontrivial interplay between the effective dimer attraction and the exchange anisotropy. 
In the easy-plane regime, the four featureless phases and two critical phases 
are found to compete in a complex manner depending on the signs of $\Jpd$. 
\end{abstract}
\maketitle


\section{Introduction}

The spin-$\frac12$ Heisenberg antiferromagnet on the square lattice 
with competing nearest-neighbor and next-nearest-neighbor interactions, $J_1$ and $J_2$, 
has attracted considerable attention as a prototypical model of a highly frustrated quantum magnet. 
While there has been a consensus on the existence of the semiclassical staggered and stripe N\'eel phases for $J_2/J_1\lesssim 0.4$ and $J_2/J_1\gtrsim 0.6$, respectively, 
the nature of the nonmagnetic ground state between the two magnetic phases has long been under debate. 
There have been different proposals including a columnar dimer (CD) \cite{PhysRevLett.66.1773,PhysRevB.60.7278, PhysRevB.97.174408} 
or plaquette \cite{PhysRevLett.84.3173, PhysRevLett.113.027201} valance bond solid (VBS) 
and a gapless \cite{PhysRevB.88.060402} or gapped \cite{PhysRevB.86.024424} quantum spin liquid (QSL).
Recent numerical studies have also suggested that 
this nonmagnetic regime consists of two different phases: a QSL and a VBS 
\cite{doi:10.7566/JPSJ.84.024720, PhysRevLett.121.107202, PhysRevB.102.014417, PhysRevX.11.031034, LIU20221034, PhysRevB.104.045110}. 


\begin{figure}
\includegraphics[width=0.4\textwidth]{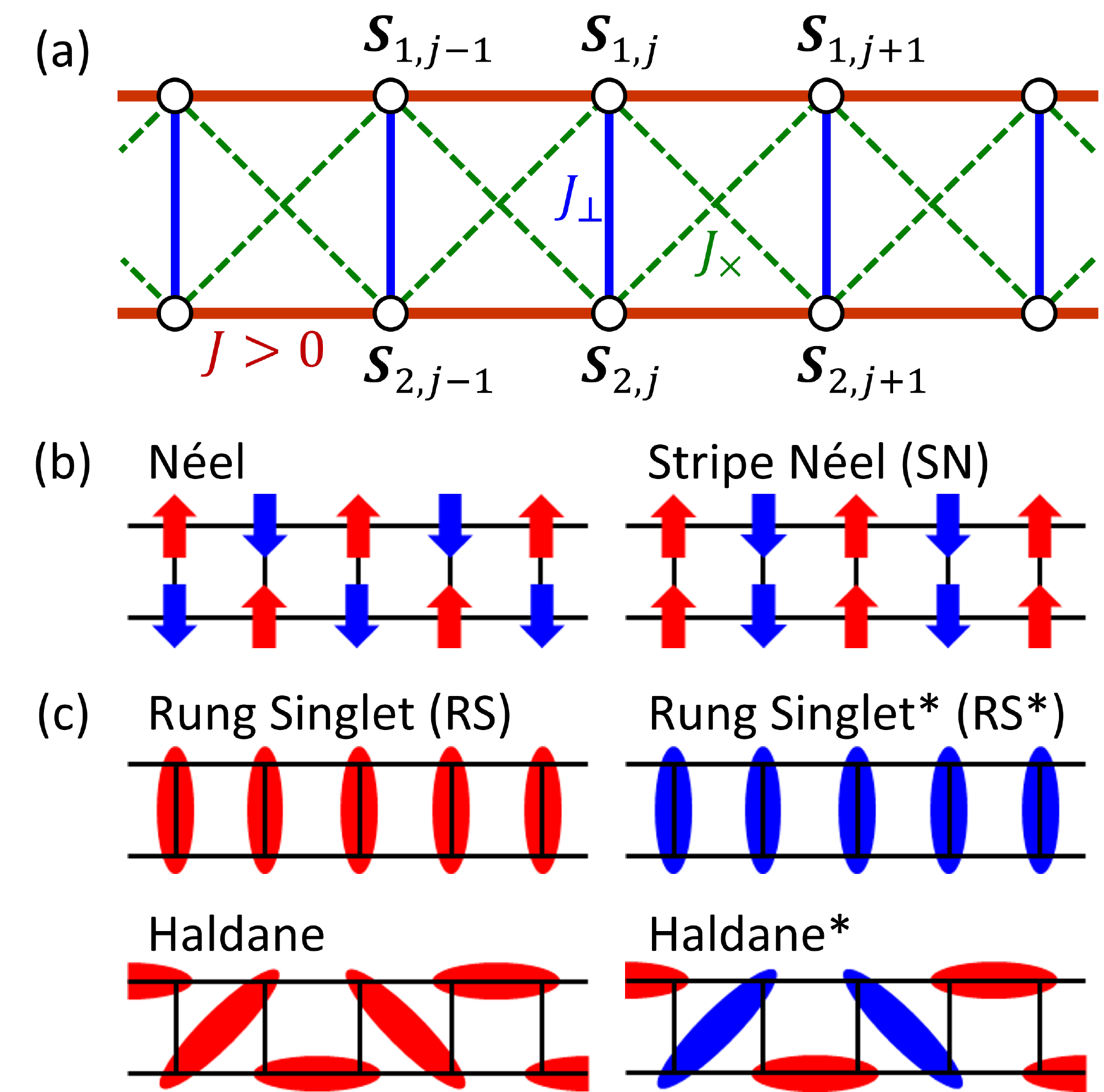} 
\caption{\label{fig:ladder_Jpd}
(a) Frustrated XXZ ladder described by the Hamiltonian \eqref{eq:ladfrust_XXZ}. 
The spin-$\frac12$ operator at the $j$th site on the $n$th leg is denoted by $\Sv_{n,j}$. 
(b) Conventional (staggered) N\'eel and stripe N\'eel (SN) states. 
(c) Four gapped featureless states with no symmetry breaking. 
A red oval indicates a singlet pair $(\ket{\uparrow\downarrow}-\ket{\downarrow\uparrow})/\sqrt{2}$. 
The Haldane state is a superposition of various singlet-covering states. 
The RS* and Haldane* states are related to the RS and Haldane states 
under the $\pi$ rotation of spins on the first leg about the $z$-axis, $U_1^z$, 
and therefore have twisted singlet pairs 
$(\ket{\uparrow\downarrow}+\ket{\downarrow\uparrow})/\sqrt{2}$ (blue oval) 
between the two legs. 
}
\end{figure}

In view of the enigmatic nature of the $J_1$-$J_2$ model in the frustrated regime $J_1\approx 2J_2$, 
it is interesting to investigate its (quasi-)one-dimensional variants 
such as a spatially anisotropic square-lattice model \cite{PhysRevB.67.024422, PhysRevLett.93.127202} 
and a ladder model \cite{PhysRevLett.93.127202, PhysRevB.81.064432},  
where powerful field-theoretical methods in one dimension can be applied. 
These models consist of an array of antiferromagnetic chains (with an intrachain coupling $J>0$), 
and neighboring chains are coupled by transverse and diagonal interactions, $\Jperp$ and $\Jdiag$. 
Based on a field-theoretical analysis for $|\Jpd|\ll J$, 
Starykh and Balents have proposed that while the phase diagram of the anisotropic square-lattice model is largely covered 
by the staggered ($\Jperp\gtrsim 2\Jdiag$) and stripe ($\Jperp\lesssim 2\Jdiag$) N\'eel phases, 
a narrow CD phase appears between these magnetic phases \cite{PhysRevLett.93.127202}.
This CD phase results from a weak effective dimer attraction between neighboring chains 
that is generated in the renormalization group (RG) process. 
This phase may continue up to $\Jperp=J$ \cite{PhysRevB.69.094418}, 
which corresponds to the square-lattice $J_1$-$J_2$ model. 
For a two-leg ladder [Fig.\ \ref{fig:ladder_Jpd}(a)], a similar analysis has predicted that a CD phase 
appears between the rung-singlet (RS) and Haldane phases \cite{PhysRevLett.93.127202, PhysRevB.81.064432}, 
as we will review in Sec.\ \ref{sec:bos_NA}. 

\begin{figure*}
\includegraphics[width=0.8\textwidth]{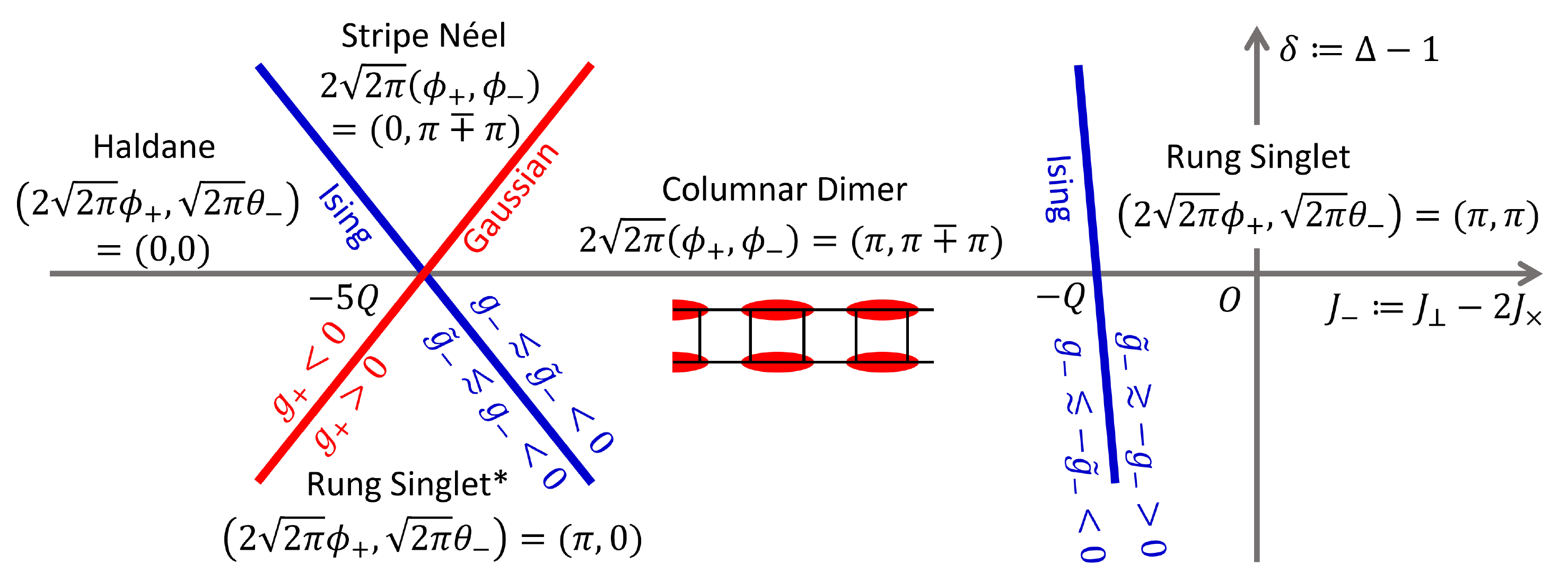} 
\caption{\label{fig:phases_Delta1}
Schematic phase diagram of the model \eqref{eq:ladfrust_XXZ} for $|\Jperp-2\Jdiag|\ll J$ and $\Delta=1+\delta$ with $|\delta|\ll 1$, 
predicted by the effective field theory. 
Here, we fix $\Jdiag\ne 0$ and vary $\Jperp$ and $\delta$. 
There are a Gaussian transition line (red) with Eq.\ \eqref{eq:JJ_del_Gauss} 
as well as two Ising transition lines (blue) with Eq.\ \eqref{eq:JJ_del_Ising}, 
where $Q$ is given by Eq.\ \eqref{eq:Q_Jx}. 
There are distinct gapped phases separated by these lines. 
Each phase is characterized by the locking positions of the bosonic fields; for more details, see Refs.\ \cite{ogino2020continuous,ogino2021spt}.  
The phase diagram for $\delta=0$ has been obtained previously in Refs.\ \cite{PhysRevLett.93.127202, PhysRevB.81.064432}. 
}
\end{figure*}

The proposal of Starykh and Balents \cite{PhysRevLett.93.127202} has stimulated numerical studies on the frustrated Heisenberg ladder model 
\cite{PhysRevB.73.224433, PhysRevB.77.205121, PhysRevB.77.214418, PhysRevB.81.064432, PhysRevB.86.075133}, 
for which large-scale simulations based on the density-matrix renormalization group (DMRG) \cite{PhysRevLett.69.2863,PhysRevB.48.10345} are feasible. 
For antiferromagnetic $\Jpd>0$, however, the presence of the CD phase has been elusive. 
By reexamining the RG analysis, Hikihara and Starykh have pointed out that 
a marginally relevant inter-leg current interaction also plays a significant role away from the weak-coupling limit $\Jpd/J\to 0$, 
and argued that a direct first-order transition between the RS and Haldane phases occurs for $\Jdiag\gtrsim 0.3J$ \cite{PhysRevB.81.064432}, 
in consistency with some numerical works \cite{PhysRevB.73.224433, PhysRevB.77.205121, doi:10.1142/S0217984900000458}. 
They have also argued that changing the signs of inter-leg interactions to ferromagnetic ($\Jpd<0$) can eliminate the issue of the marginal interaction altogether, 
and they have numerically demonstrated that the CD phase appears over a wide region around $\Jperp\approx 2\Jdiag<0$.  
This is a remarkable example of a highly frustrated quantum magnet 
in which a nontrivial consequence of a RG-generated interaction is clearly confirmed in a numerical simulation 
(see, e.g., Refs.\ \cite{PhysRevB.94.035154, PhysRevLett.98.077205, PhysRevB.84.174415, 2205.15525} for other examples). 
Furthermore, this system may share similar frustration physics with the long-debated regime of the $J_1$-$J_2$ antiferromagnet on the square lattice. 

Motivated by the intriguing ground-state properties of the spin-$\frac12$ frustrated Heisenberg ladder, 
in this paper, we study an XXZ extension of the model [Fig.\ \ref{fig:ladder_Jpd}(a)]. 
The Hamiltonian of our model is given by 
\begin{align}\label{eq:ladfrust_XXZ}
H  =& J \sum_{n= 1}^2 \sum_{j}  (\Sv_{n,j} \cdot \Sv_{n,j+1})_\Delta
	 + \Jperp \sum_j (\Sv_{1,j}\cdot\Sv_{2,j})_\Delta \notag\\
   &+ J_\times \sum_{j} \left[ (\Sv_{1,j}\cdot\bm{S}_{2,j+1})_\Delta+(\Sv_{2,j}\cdot\bm{S}_{1,j+1})_\Delta \right],
\end{align}
where
\begin{equation}\label{eq:XXZ}
 (\Sv_{n,i}\cdot\Sv_{m,j})_\Delta := S^x_{n,i} S^x_{m,j} + S^y_{n,i} S^y_{m,j} + \Delta S^z_{n,i} S^z_{m,j} .
\end{equation}
We set $J=1$ as the unit of energy when presenting numerical results. 
An XXZ anisotropy enriches the phase structure in the following way. 
An easy-axis anisotropy $\Delta>1$ tends to stabilize semiclassical collinear magnetic orders along the $z$-axis. 
We therefore expect the conventional (staggered) N\'eel phase with $\langle S_{n,j}^z\rangle\propto (-1)^{n+j}$ for $\Jperp\gtrsim 2\Jdiag$ 
and the stripe N\'eel (SN) phase with $\langle S_{n,j}^z\rangle\propto (-1)^j$ for $\Jperp\lesssim 2\Jdiag$; see Fig.\ \ref{fig:ladder_Jpd}(b). 
Meanwhile, an easy-plane anisotropy $\Delta<1$ allows for the emergence of two additional gapped featureless phases with no symmetry breaking: 
the RS* and Haldane* phases \cite{PhysRevB.86.195122, Fuji15, PhysRevB.93.104425, PhysRevB.87.081106, Li_2017, ogino2021spt}. 
These are related to the RS and Haldane phases under the $\pi$ rotation of spins about the $z$-axis on the first leg, $U_1^z:={\exp} ( i\pi \sum_j S_{1,j}^z )$, 
and therefore have twisted singlet pairs $(\ket{\uparrow\downarrow}+\ket{\downarrow\uparrow})/\sqrt{2}$ between the two legs 
\footnote{
The RS, RS*, Haldane, and Haldane* phases are called the rung-$\ket{0,0}$, rung-$\ket{1,z}$, $t_0$, and $t_z$ phases in Ref.\ \cite{PhysRevB.86.195122}. 
}; 
see Fig.\ \ref{fig:ladder_Jpd}(c). 
The RS and Haldane phases and their twisted variants are all distinct 
in the presence of the $D_2\times\sigma$ and translational symmetries \cite{PhysRevB.86.195122, Fuji15, ogino2021spt}, 
where $D_2=\mathbb{Z}_2\times\mathbb{Z}_2$ is the dihedral group of $\pi$ spin rotations and $\sigma$ is the symmetry with respect to the interchange of the two legs. 

We are particularly interested in the interplay between the RG-generated dimer attraction and the XXZ anisotropy around the isotropic case $\Delta=1$. 
An insight into this problem can be gained from previous works on an XXZ ladder 
with an {\it explicit} inter-leg dimer attraction term \cite{PhysRevLett.78.3939, PhysRevB.82.214420, PhysRevLett.122.027201, ogino2021spt}. 
By combining the bosonization analyses of Refs.\ \cite{PhysRevLett.93.127202, PhysRevB.81.064432, PhysRevB.82.214420, ogino2021spt}, 
we argue that the Haldane-CD transition point in the isotropic model is a crossing point of two transition lines in the XXZ model, 
and that the SN and RS* phases appear between these lines, 
as shown in Fig.\ \ref{fig:phases_Delta1}. 
Namely, the RG-generated dimer attraction not only induces the CD phase 
but also brings about a rich phase structure around it in the presence of the exchange anisotropy. 

\begin{figure}
\includegraphics[width=0.45\textwidth]{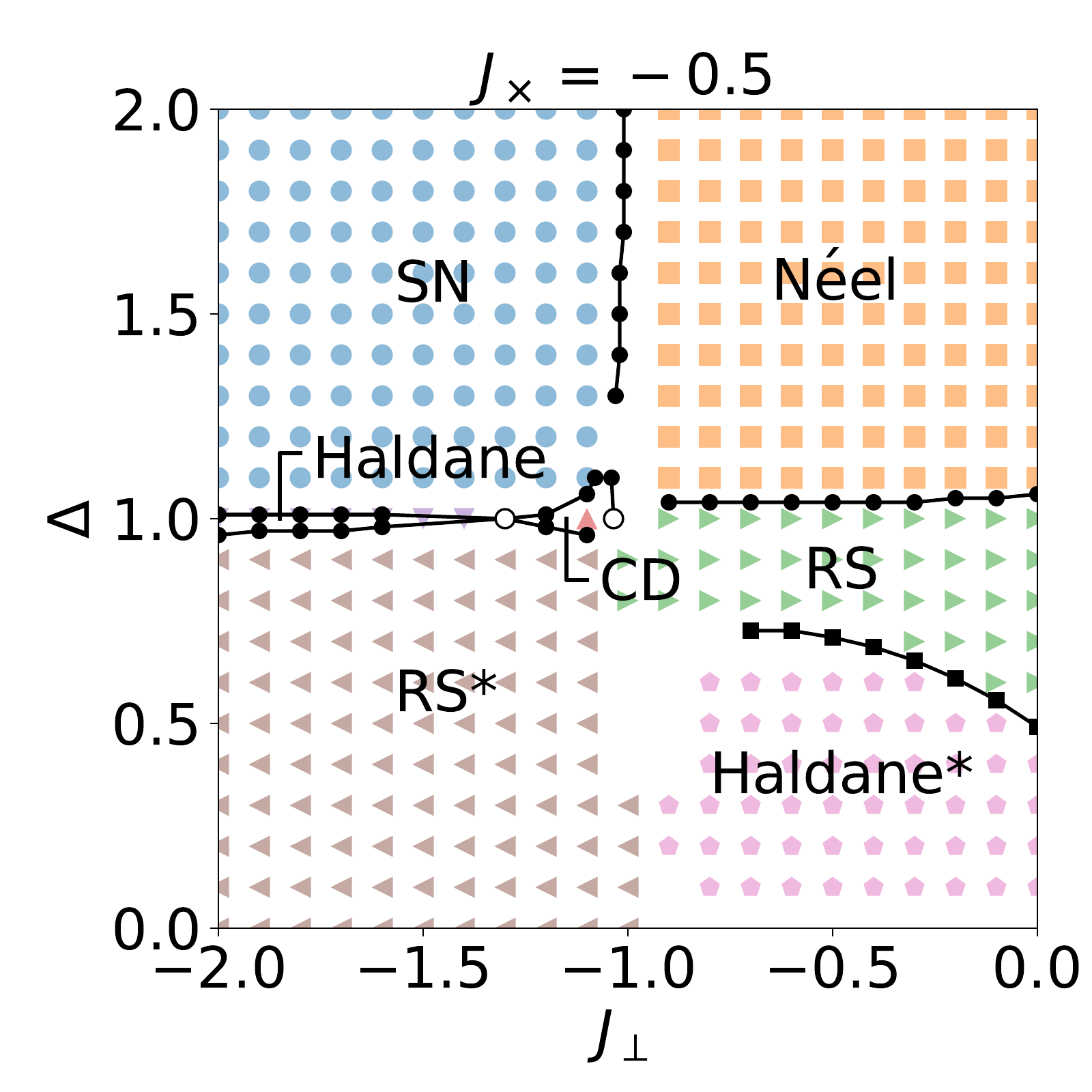}
\caption{\label{fig:phasediagram}
Phase diagram of the model \eqref{eq:ladfrust_XXZ} with $\Jdiag = -0.5$, obtained numerically. 
We set $J=1$ as the unit of energy. 
Symmetry-broken phases (N\'eel, SN, and CD) are identified by calculating the corresponding order parameters (Sec.\ \ref{sec:Orders}). 
Featureless phases (RS, RS*, Haldane, and Haldane*) are identified by calculating topological indices (Sec.\ \ref{sec:topo_indices}). 
The Haldane*-RS transition points (black squares) are determined by the level spectroscopy analysis (Appendix \ref{app:num_data}). 
The Haldane-CD-RS transition points for $\Delta=1$ (empty circles) are taken from the DMRG results of Ref.\ \cite{PhysRevB.81.064432}. 
Other transition points (black circles) are determined by the iDMRG analysis of the order parameters and the correlation length. 
In some region near $\Jperp= -1$, the phase structure 
could not be determined 
because of an insufficient convergence of the iDMRG. 
}
\end{figure}

We determine the ground-state phase diagram over a wide range of $\Delta$ and for different signs of $\Jpd$ 
by means of the Abelian bosonization, 
the infinite DMRG (iDMRG) method \cite{McCulloch_2007, 0804.2509, 10.21468/SciPostPhysLectNotes.5}, 
and numerical exact diagonalization \cite{SciPostPhys.2.1.003, 10.21468/SciPostPhys.7.2.020}. 
The phase diagram obtained numerically for $\Jdiag/J=-0.5$ in Fig.\ \ref{fig:phasediagram} 
confirms the phase structure near $\Delta=1$ in Fig.\ \ref{fig:phases_Delta1}. 
This phase diagram is based mainly on the iDMRG calculations of the correlation length, order parameters, and topological indices. 
We also find that in the easy-plane regime $-1<\Delta<1$, 
the four gapped featureless phases and two critical phases (XY1 and XY2) 
compete in a complex manner depending on the signs of $\Jpd$ 
[Figs.\ \ref{fig:phases_Delta0}(b,c) and \ref{fig:phasediagram_Jxpm0.1} shown later]. 
The transition between a critical phase and a featureless phase is 
of the Berezinskii-Kosterlitz-Thouless (BKT) type \cite{Berezinsky:1970fr, Berezinsky:1972rfj, Kosterlitz_1973,Kosterlitz_1974}
while the RS*-Haldane and Haldane*-RS transitions are of the Gaussian type. 
Both of these transitions can be described by an effective sine-Gordon theory. 
We determine the BKT and Gaussian transition lines accurately by means of the level spectroscopy method
\cite{Nomura_1995, Kitazawa_1997, Nomura_1998, PhysRevB.59.11358, cond-mat/0201072, PhysRevB.72.014449, PhysRevB.104.165132}, 
which combines the sine-Gordon theory with exact diagonalization. 
With this method, we have also obtained an accurate phase diagram of the simple XXZ ladder model with $\Jdiag=0$ 
(Fig.\ \ref{fig:phasediagram_j_jp_model} shown later); 
this phase diagram is improved from the one obtained earlier by Li {\it et al.} \cite{Li_2017} 
in that the former is now consistent with the Abelian bosonization analysis [Fig.\ \ref{fig:phases_Delta0}(a)]. 
We note that the iDMRG and the level spectroscopy employed in the present work have different advantages and complement each other: 
the iDMRG is most efficient in addressing the properties of gapped phases in the thermodynamic limit 
while the level spectroscopy can accurately determine certain transition lines by relying on the knowledge of the underlying field theory. 

The rest of this paper is organized as follows. 
In Sec.\ \ref{sec:EFT}, we describe effective field theory based on bosonization, 
and discuss qualitative features of the ground-state phase diagram. 
In Sec.\ \ref{sec:iDMRG}, we present numerical results obtained by the iDMRG method. 
In particular, we analyze order parameters and topological indices, which can detect symmetry-broken phases and featureless phases, respectively. 
In Sec.\ \ref{sec:LevelSpec}, we describe the level spectroscopy analysis, 
and present detailed phase diagrams in the easy-plane regime. 
In Sec.\ \ref{sec:summary}. we present a summary of the present study and an outlook for future studies. 
In Appendix \ref{app:spectra_eff}, we describe simple perturbative calculations of the finite-size spectra of the effective Hamiltonians. 
These calculations reproduce some of the previous results 
\cite{Nomura_1995, Kitazawa_1997, Nomura_1998, PhysRevB.59.11358, cond-mat/0201072, PhysRevB.72.014449, PhysRevB.104.165132}, 
and also give the theoretical basis of our level spectroscopy analysis. 
In Appendix \ref{app:num_data}, we provide some supplemental numerical data. 

\section{Effective field theory}\label{sec:EFT}

For weak inter-chain couplings with $|\Jpd | \ll J$, the ground-state phase diagram of the model \eqref{eq:ladfrust_XXZ} can be studied 
by means of effective field theory based on bosonization \cite{giamarchi2003quantum,gogolin2004bosonization}. 
After reviewing previous results on the isotropic model \cite{PhysRevLett.93.127202,PhysRevB.81.064432} in Sec.\ \ref{sec:bos_NA}, 
we analyze the XXZ model by means of the Abelian bosonization in Sec.\ \ref{sec:bos_A}. 

\subsection{Non-Abelian bosonization for the Heisenberg ladder}\label{sec:bos_NA}

We first review the non-Abelian bosonization analysis of the frustrated Heisenberg ladder \eqref{eq:ladfrust_XXZ} 
with $\Delta=1$ \cite{PhysRevLett.93.127202,PhysRevB.81.064432,PhysRevB.77.205121,PhysRevB.95.144415}. 

In the limit $\Jpd/J\to 0$, each isolated antiferromagnetic Heisenberg chain is described 
by the SU(2)$_1$ Wess-Zumino-Novikov-Witten (WZNW) theory (with the spin velocity $v=\pi Ja/2$) 
perturbed by a marginally irrelevant backscattering term 
\cite{Affleck88_bos, gogolin2004bosonization, DiFrancesco97, PhysRevB.54.R9612}. 
The spin operators on the $n$th leg ($n=1,2$) can be decomposed as
\begin{equation}
 \Sv_{n,j} \to a [\Mv_n (x) + (-1)^j \Nv_n(x)]
\end{equation}
with $x=ja$, where $a$ is the lattice spacing of each chain.
The uniform and staggered components, $\Mv_n$ and $\Nv_n$,
have the scaling dimensions $1$ and $1/2$, respectively.
The former can further be decomposed
into chiral (right and left) components as
$\Mv_n=\Mv_{nR}+\Mv_{nL}$.  Another important operator is the
(in-chain) staggered dimerization operator $\epsilon_n$ defined by
\begin{equation}
   (-1)^j \Sv_{n,j}\cdot\Sv_{n,j+1} \to a \epsilon_n (x),  
\end{equation}
which has the scaling dimension $1/2$. 

The interchain couplings $\Jpd$ produce various symmetry-allowed perturbations around the WZNW fixed point. 
Such perturbations are summarized as 
\begin{equation}\label{eq:Hp_NA}
 H' = \int dx \sum_i g_i \Ocal_i,
\end{equation}
where $i$ runs over the following 6 operators: 
\begin{subequations}\label{eq:Hp_NA_terms}
\begin{align}
 \Ocal_\bs &= \Mv_{1R} \cdot \Mv_{1L} + \Mv_{2R} \cdot \Mv_{2L},\\
 \Ocal_1 &= \Mv_{1R} \cdot \Mv_{2L} + \Mv_{1L} \cdot \Mv_{2R},\\
 \Ocal_2 &= \Mv_{1R} \cdot \Mv_{2R} + \Mv_{1L} \cdot \Mv_{2L},\\
 \Ocal_N &= \Nv_1\cdot\Nv_2,\\
 \Ocal_\epsilon &= \epsilon_1 \epsilon_2,\\
 \Ocal_{\partial N} &= a^2 \partial_x \Nv_1\cdot \partial_x\Nv_2. 
\end{align}
\end{subequations}
Here, we have neglected terms with higher scaling dimensions. 
The bare coupling constants are given by 
\begin{equation}\label{eq:g_bare}
\begin{split}
 &g_1(0)=g_2(0)=(\Jperp+2\Jdiag) a,\\
 &g_N(0)=(\Jperp-2\Jdiag)a,~g_\epsilon(0)=0,\\
 &g_{\partial N}(0)=\Jdiag a,
\end{split}
\end{equation}
and $g_\bs(0)$ is estimated in Ref.\ \cite{PhysRevB.54.R9612}. 
We may ignore the coupling $g_2$ in the following argument 
as its evolution in the RG flow is decoupled from the other couplings. 

Among the perturbations in Eq.\ \eqref{eq:Hp_NA}, 
the $g_N$ and $g_\epsilon$ terms with the scaling dimension $1$ are relevant and grow exponentially in the RG flow. 
As the initial value of $g_\epsilon$ is zero, the $g_N$ term dominantly determines the phase structure almost everywhere in the parameter space $(\Jperp,\Jdiag)$. 
This term induces the RS and Haldane phases for $\Jperp\gtrsim 2\Jdiag$ and $\Jperp\lesssim 2\Jdiag$, respectively. 
Near the line $\Jperp=2\Jdiag$, however, a more careful consideration is required 
as $g_N(0)$ also vanishes on this line. 
It has been argued that in this regime, $g_\epsilon$ acquires a negative value of order $\Jdiag^2/J$ 
through the fusion of the $g_1$ and $g_{\partial N}$ terms in an early stage of the RG flow 
\cite{PhysRevLett.93.127202,PhysRevB.81.064432}. 
Similarly, the $g_N$ coupling acquires a positive correction. 
These corrections can be taken into account by renormalizing the initial values as
\begin{subequations}\label{eq:gNE_initial}
\begin{align}
 g_N(0)&\to (\Jperp-2\Jdiag+2Q)a ,\\
 g_\epsilon(0)&\to -3Qa,
\end{align}
\end{subequations}
where
\begin{equation}\label{eq:Q_Jx}
Q=\frac{g_1(0)g_{\partial N}(0)}{4a (2\pi v)}\approx \frac{\Jdiag^2}{\pi^2 J}.
\end{equation}
Using these renormalized values, the competition between the $g_N$ and $g_\epsilon$ terms 
has been analyzed through the refermionization procedure \cite{PhysRevLett.93.127202,PhysRevB.81.064432}.  
This analysis has predicted the emergence of the CD phase in a narrow region 
\begin{equation}\label{eq:CD_isotropic}
-5Q<\Jperp-2\Jdiag<-Q.
\end{equation} 
See the $\delta=0$ case of Fig.\ \ref{fig:phases_Delta1}. 
We extend this result to the XXZ case by using the Abelian bosonization in Sec.\ \ref{sec:bos_A_Delta1}. 

In the above argument, we have focused on the relevant couplings $g_N$ and $g_\epsilon$, 
and neglected the roles of the marginal couplings, especially, $g_1$ in the long-distance physics. 
The focus on the relevant couplings is justified asymptotically in the limit $\Jpd/J\to 0$. 
Away from this limit, however, one has to take account of the marginally relevant nature of $g_1$ for antiferromagnetic $\Jpd>0$. 
By numerically solving the RG equations, it has been found that for $\Jdiag\gtrsim 0.3J$ and near the line $\Jperp=2\Jdiag$, 
the coupling $g_1$ reaches the order of $J$ earlier than $g_N$ and $g_\epsilon$ \cite{PhysRevB.81.064432}. 
The dominance of the $g_1>0$ term results in a direct first-order phase transition between the RS and Haldane phases 
\footnote{
This can be seen in the effective Hamiltonian \eqref{eq:Heff} in the Abelian bosonization. 
If only the $\gamma_1>0$ term is considered as a perturbation, 
it results in the two degenerate ground states $(2\sqtp\phi_+,\sqtp\theta_-)=(\pi,\pi),(0,0)$, 
which correspond to the RS and Haldane phases, respectively. 
One of these phases are selected by infinitesimal $g_+\ne 0$ or $\tilde{g}_-\ne 0$. 
See Refs.\ \cite{PhysRevLett.93.127202,PhysRevB.81.064432} for more detail. 
A similar first-order transition also occurs on a zigzag ladder with bond alternation; see Fig.\ 11 of Ref.\ \cite{PhysRevB.86.094417}. 
}. 
Therefore, to obtain the CD phase in the antiferromagnetic model, 
one is expected to focus on the limited region $0<\Jdiag\lesssim 0.3J$. 
The situation significantly changes for ferromagnetic $\Jpd<0$. 
In this case, $g_1$ is marginally irrelevant, 
and it is legitimate to focus on the relevant couplings $g_N$ and $g_\epsilon$ in the present perturbative argument. 
It has been numerically demonstrated in Ref.\ \cite{PhysRevB.81.064432} 
that the CD phase indeed appears over an extended region for $\Jpd<0$; 
in consistency with Eq.\ \eqref{eq:CD_isotropic}, the range of $\Jperp-2\Jdiag$ in which the CD order appears 
expands with an increase in $|\Jdiag|$. 

\subsection{Abelian bosonization for the XXZ ladder}\label{sec:bos_A}

We now consider the XXZ ladder \eqref{eq:ladfrust_XXZ} using the Abelian bosonization formalism. 
This approach has successfully been used for related problems of spin-$\frac12$ ladders 
\cite{PhysRevLett.69.2419,PhysRevB.50.9911,PhysRevB.53.8521,PhysRevLett.78.3939,PhysRevB.73.214427, PhysRevB.62.14965, 
PhysRevB.82.214420,PhysRevB.66.134423,ogino2020continuous,ogino2021spt} 
and zigzag ladders \cite{PhysRevLett.81.910, PhysRevB.62.14965, PhysRevB.86.094417}. 
Here we take similar notations as those in Refs.\ \cite{PhysRevB.82.214420,ogino2020continuous,ogino2021spt} 
so that we can simplify the explanation of the formalism. 

In the limit $\Jpd/J \to 0$, each chain labeled by $n=1,2$ is described 
by the Tomonaga-Luttinger liquid (TLL) theory (the Gaussian theory) 
in terms of a dual pair of bosonic fields $\phi_n(x)$ and $\theta_n(x)$. 
The velocity $v$ and the TLL parameter $K$ are known from the exact solution as 
\begin{align}\label{eq:vK_XXZ}
 v=\frac{\sin (\pi \eta)}{2(1-\eta)} Ja,~~
 K=\frac{1}{2\eta},
\end{align}
where the anisotropy is parameterized as $\Delta=-\cos(\pi\eta)~(0<\eta\le 1)$. 
The spin operators are related to the bosonic fields as 
\begin{subequations}\label{eq:Spin_dim_bos}
\begin{align}
&S_{n, j}^z = \frac{a}{\sqrt{\pi}} \partial_x \phi_n + (-1)^j a_1 \cos(2\sqrt{\pi} \phi_n) + \cdots, \label{eq:Sz_bos}\\
&S_{n,j}^+ = e^{i\sqrt{\pi}\theta_n} \qty[b_0 (-1)^j + b_1 \cos(2\sqrt{\pi}\phi_n) + \cdots],\label{eq:Sp_bos}
\end{align}
\end{subequations}
where $S_{n,j}^\pm:=S_{n,j}^x\pm iS_{n,j}^y$, and $a_1$, $b_0$, and $b_1$ are non-universal coefficients 
\cite{LUKYANOV1997571,PhysRevB.58.R583}. 
For $\Delta=1$, these coefficients satisfy $a_1 = b_0 =: \bar{a}$ because of the SU$(2)$ symmetry. 

Treating $\Jpd$ perturbatively, we obtain the low-energy effective Hamiltonian of the model (\ref{eq:ladfrust_XXZ}) as 
\begin{equation}\label{eq:Heff}
\begin{split}
H^\mathrm{eff}
&=\int \dd x \sum_{\nu=\pm} \frac{v_\nu}{2} \qty[ \frac1{K_\nu} \qty(\partial_x\phi_\nu) ^2+ K_\nu \qty(\partial_x\theta_\nu)^2] \\
&+ g_+ \cos(2\sqrt{2\pi}\phi_+) + g_- \cos(2\sqrt{2\pi}\phi_-) \\
&+ \tilde{g}_- \cos(\sqrt{2\pi}\theta_-) - \gamma_1 \cos(2\sqtp\phi_+) \cos(\sqtp\theta_-) \\
&+ \gamma_\mathrm{bs} \cos(2\sqrt{2\pi}\phi_+)\cos(2\sqrt{2\pi}\phi_-) +\dots,
\end{split}
\end{equation}
where
\begin{equation}
\phi_{\pm} := \frac{1}{\sqrt{2}} (\phi_1 \pm \phi_2), ~~
\theta_{\pm} := \frac{1}{\sqrt{2}} (\theta_1 \pm \theta_2), 
\end{equation}
and ellipses indicate terms that have higher scaling dimensions. 
For $\Delta=1$, the coupling constants $g_\pm$, $\tilde{g}_-$, $\gamma_1$, and $\gamma_\bs$ 
are related to $g_N\mp g_\epsilon$, $g_N$, $g_1$, and $g_\bs$, respectively, in Eq.\ \eqref{eq:Hp_NA}  
\footnote{See Ref.\ \cite{PhysRevB.86.094417} for the expressions of the operators in Eq.\ \eqref{eq:Hp_NA_terms} in the Abelian bosonization. 
Here, $\phi_\pm$ and $\theta_\pm$ in Ref.\ \cite{PhysRevB.86.094417} correspond to $\sqrt{2}\phi_\pm$ and $\theta_\pm/\sqrt{2}$ in the present paper}. 
The bare coupling constants are given by  
\begin{subequations}\label{eq:coeff_cos}
\begin{align}
 g_\pm &= \frac{a_1^2}{2a} \qty[ \qty(J_-+2Q)\Delta \pm 3Q], \label{eq:coeff_cos_gpm}\\
 \tilde{g}_-&=\frac{b_0^2}{a}  (J_-+2Q), \label{eq:coeff_gtm}\\
 \gamma_1&= 
 \frac{b_1^2}{2a} J_+,
\end{align}
\end{subequations}
and $\gamma_\bs$ is given in Refs.\ \cite{LUKYANOV1998533,LUKYANOV2003323}. 
Here, we take the shorthand notation $J_\pm:=\Jperp\pm 2\Jdiag$. 
Furthermore, we have taken into account the correction $Q$ in Eq.\ \eqref{eq:gNE_initial} that appears for $\Delta=1$. 
While the correction $Q$ has not been estimated for $\Delta\ne 1$, 
we may take the value at $\Delta=1$ in Eq.\ \eqref{eq:Q_Jx} if $\Delta$ is sufficiently close to 1. 
If $\Delta$ is away from 1, we neglect the correction $Q$ 
as the CD phase induced by the correction is not found for such $\Delta$ in the numerical results; see Fig.\ \ref{fig:phasediagram}.  
The velocities $v_\pm$ and the TLL parameters $K_\pm$ in the symmetric and antisymmetric channels 
are modified from $v$ and $K$ in the decoupled XXZ chains [Eq.\ \eqref{eq:vK_XXZ}] by the effects of the inter-chain couplings as 
\begin{subequations}\label{eq:vpm_Kpm}
\begin{align}
 v_\pm&=v\left(1\pm \frac{aK}{\pi v} J_+ \Delta \right)^{1/2},\\
 K_\pm&=K\left(1\pm \frac{aK}{\pi v} J_+ \Delta \right)^{-1/2}. \label{eq:Kpm}
\end{align}
\end{subequations} 

\subsubsection{Case of $\Delta\approx 1$}\label{sec:bos_A_Delta1}

As explained in Sec.\ \ref{sec:bos_NA}, the competition between the $g_N$ and $g_\epsilon$ terms 
results in the emergence of the CD phase in a narrow region in Eq.\ \eqref{eq:CD_isotropic} for $\Delta=1$. 
We now extend this result to the weakly anisotropic regime, i.e.,  $\Delta=1+\delta$ with $|\delta|\ll 1$, by applying the Abelian bosonization formalism. 
To this end, we focus on the Gaussian part as well as the relevant $g_\pm$ and $\tilde{g}_-$ terms in the effective Hamiltonian \eqref{eq:Heff}. 
We note that the $g_\pm$ and $\tilde{g}_-$ terms have the scaling dimensions $2K_\pm$ and $(2K_-)^{-1}$, 
respectively, which are all equal to unity in the limit of decoupled Heisenberg chains where $K_\pm=1/2$. 
Neglect of the other terms such as $\gamma_1$ is justified asymptotically in the limit $\Jpd/J \to 0$ 
as they have higher scaling dimensions. 
We expect that even away from this limit, 
the prediction of the present analysis would hold qualitatively 
as long as the CD phase exists over a finite interval between the RS and Haldane phases in the isotropic case $\Delta=1$. 

The remaining analysis can be done in the same way as 
our previous studies on the XXZ ladder with an explicit inter-leg dimer interaction \cite{ogino2020continuous,ogino2021spt}. 
As we have neglected $\gamma_1$, $\gamma_\bs$, and the terms with higher scaling dimensions, 
the effective Hamiltonian \eqref{eq:Heff} is decoupled into the symmetric and antisymmetric channels. 
The symmetric channel is described by the sine-Gordon model, 
in which the strongly relevant $g_+$ term locks $\phi_+$ at distinct positions depending on the sign of $g_+$. 
A Gaussian transition with the central charge $c=1$ is expected at $g_+=0$. 
The antisymmetric channel is described by the dual-field double sine-Gordon model, 
in which the strongly relevant $g_- $ and $\tilde{g}_-$ terms compete. 
As these terms have close scaling dimensions (equal in the limit of decoupled Heisenberg chains), 
the long-distance physics can be determined by examining which of $|g_-|$ and $|\tilde{g}_-|$ is larger. 
Specifically, $|g_-|\gtrsim |\tilde{g}_-|$ ($|g_-|\lesssim |\tilde{g}_-|$) leads to the locking of $\phi_-$ ($\theta_-$), 
and an Ising transition with the central charge $c=1/2$ is expected at $|g_-|\approx |\tilde{g}_-|$ \cite{PhysRevB.53.8521,LECHEMINANT2002502}. 

Based on the above analysis and using the bare coupling constants in Eq.\ \eqref{eq:coeff_cos}, 
we obtain the schematic phase diagram in Fig.\ \ref{fig:phases_Delta1}. 
In this diagram, there are a Gaussian transition line with
\begin{equation}\label{eq:JJ_del_Gauss}
 J_- =-(5-3\delta)Q
\end{equation}
as well as two Ising transition lines with
\begin{subequations}\label{eq:JJ_del_Ising}
\begin{align}
J_- &\approx -(5+3\delta)Q, \label{eq:JJ_del_Ising1}\\
J_- &\approx -(1+\delta/3)Q. 
\end{align}
\end{subequations}
The Haldane-CD transition point in the isotropic model is now a crossing point of 
the two transitions lines with Eqs.\ \eqref{eq:JJ_del_Gauss} and \eqref{eq:JJ_del_Ising1}. 
Just with an infinitesimal anisotropy $\delta\ne0$, the SN and RS* phases appear around this point. 
It is worth emphasizing that this rich phase structure is provided 
by the RG-generated dimer attraction, specifically, the corrections $\pm 3Q$ in Eq.\ \eqref{eq:coeff_cos_gpm}. 
Without the corrections, only the RS and Haldane phases would appear near $\Delta=1$. 
We also note that Fig.\ \ref{fig:phases_Delta1} has a similar structure as the phase diagram of an XXZ ladder with an explicit dimer attraction term; 
see Fig. 3 of Ref.\ \cite{ogino2021spt}. 


\subsubsection{Case of easy-axis anisotropy}\label{sec:bos_A_easyaxis}

The phase structure in Fig.\ \ref{fig:phases_Delta1} holds only over a narrow range of $\Delta$ around $1$. 
In particular, the CD phase is not confirmed for $\Delta\lesssim 0.9$ or $\Delta\gtrsim 1.3$ 
in the numerically obtained phase diagram in Fig.\ \ref{fig:phasediagram}.  
We thus neglect the correction $Q$ in Eq.\ \eqref{eq:coeff_cos} for $\Delta$ away from 1, as noted earlier. 

Here we focus on the case of easy-axis anisotropy $\Delta>1$, and also assume that $\Delta$ is sufficiently away from 1, to simplify the discussion. 
In this regime, we can focus on the perturbations $g_\pm$ and $\gamma_\bs$ in the effective Hamiltonian \eqref{eq:Heff} 
as we have $K_\pm < 1/2$ and the perturbations related to $\phi_-$ are more relevant than those related to $\theta_-$. 
Specifically, $g_-$ is more relevant than $\tilde{g}_-$, and $\gamma_\bs$ is more relevant than $\gamma_1$. 
In the simplified effective Hamiltonian in which $\tilde{g}_-$ and $\gamma_1$ are neglected, 
we obtain the N\'eel and SN phases with $2\sqrt{2\pi}(\phi_+,\phi_-)=(\pi,\mp\pi)$ and $(0,\pi\mp\pi)$ for $\Jperp>2\Jdiag$ and $\Jperp<2\Jdiag$, respectively, 
and a first-order phase transition occurs between them because of the backscattering term with $\gamma_\bs<0$. 
As the effects of the neglected terms are nontrivial, the present argument does not give a strong support for a first-order transition. 
Yet, the present argument can provide an effective picture for the numerical result (presented later in Fig.\ \ref{fig:order_SN_Neel} in Sec.\ \ref{sec:Orders}) 
that indicates a first-order transition. 
We note that a first-order transition between the N\'eel and SN phases naturally occurs for $\Jdiag=0$ and $\Delta>1$ (Fig.\ \ref{fig:phasediagram_j_jp_model}) 
as the two types of states are degenerate in the decoupled-chain limit $\Jperp\to 0$. 

\subsubsection{Case of easy-plane anisotropy}\label{sec:bos_A_easyplane}

\newcommand{\Gt}{\tilde{G}}
\newcommand{\Kcal}{{\cal K}}

\begin{figure}
\includegraphics[width=0.43\textwidth]{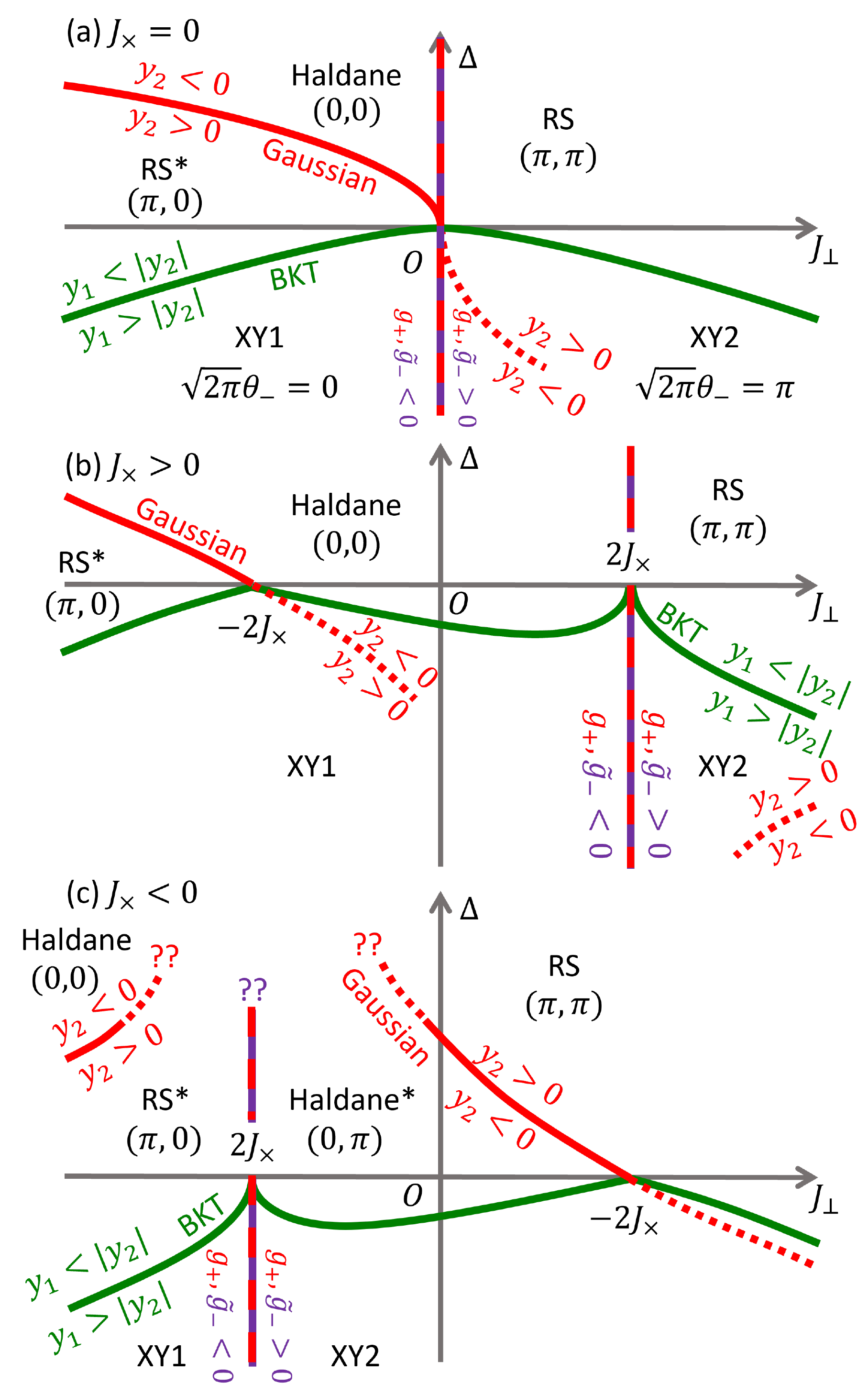} 
\caption{\label{fig:phases_Delta0}
Schematic phase diagrams of the model \eqref{eq:ladfrust_XXZ}  in the easy-plane regime 
for (a) $\Jdiag=0$, (b) $\Jdiag>0$, and (c) $\Jdiag<0$, predicted by the effective field theory. 
Here, we fix $\Jdiag$ in each case, and vary $\Jperp$ and $\Delta$. 
Two numbers for each gapped featureless phase indicate the locking positions of $(2\sqrt{2\pi}\phi_+,\sqrt{2\pi}\theta_-)$. 
Gaussian (red) and BKT (green) transition lines are given by Eqs.\ \eqref{eq:Delta_Gauss} and \eqref{eq:Delta_BKT}, respectively. 
According to the symmetry argument in Ref.\ \cite{PhysRevB.72.014449,Kitazawa_2003}, 
however, the Haldane-XY1 and Haldane*-XY2 transition lines should be located exactly on the $\Delta=0$ line. 
The present approach has a limitation in that the role of such a symmetry in the RG flow is not taken into account. 
The Gaussian lines for $\Delta<0$ (red dotted lines), which are inside the critical phases and do not correspond to transitions, are shown for reference. 
The vertical lines with alternating red and purple colors at $\Jperp=2\Jdiag$ indicate the sign changes of $g_+$ and $\gt_-$ in Eq.\ \eqref{eq:coeff_cos}; 
for $\Jdiag\ne 0$, these sign changes can occur at different points, whose description is beyond the scope of the present theory. 
See Figs.\ \ref{fig:phasediagram_j_jp_model} and \ref{fig:phasediagram_Jxpm0.1} for the phase diagrams obtained numerically. 
}
\end{figure}

Here we focus on the case of easy-plane anisotropy $-1<\Delta<1$, and also assume that $\Delta$ is sufficiently away from $\pm 1$. 
In this regime, we have $K_\pm>1/2$, and thus the $\gt_-$ term with the scaling dimension $(2K_-)^{-1}<1$ is the most relevant perturbation in the effective Hamiltonian \eqref{eq:Heff}. 
Therefore, we may assume that $\theta_-$ is locked into the minimum of this term, 
leading to a nonzero expectation value $\mu:=-\langle \cos(\sqrt{2\pi}\theta_-) \rangle\ne 0$, which has the same sign as $\gt_-$. 
We again neglect the correction $Q$ in Eq.\ \eqref{eq:coeff_cos}. 
The coupling constant $\gt_-$ grows in the RG flow as 
\begin{equation}
 \gt_-(\ell)=\gt_-(0) \exp\{ \qty[2-(2K_-)^{-1}] \ell\} 
\end{equation}
as the short-distance cutoff $\alpha$ is changed to $e^\ell \alpha$. 
We continue the RG transformation until $\gt_-(\ell)$ becomes $O(v_-)=O(J)$. 
At this scale ($\ell=\ell_1$), $\theta_-$ is locked tightly, 
and hence $\mu$ is expected to be a constant $\mu_1$ of order unity. 
We can then treat the $\gamma_1$ term in Eq.\ \eqref{eq:Heff} in a mean-field manner 
by replacing $-\cos(\sqrt{2\pi}\theta_-)$ by its mean value $\mu_1$, 
and combine the resulting expression with the $g_+$ term. 
In this way, we obtain an effective sine-Gordon model for the symmetric channel as 
\begin{equation}\label{eq:Hp_eff}
\begin{split}
H_+^\mathrm{eff}
=\int \dd x &\frac{v_+}{2} \qty[\frac{1}{K_+}\qty(\partial_x\phi_+) ^2+ K_+ \qty(\partial_x\theta_+)^2] \\
&+ g_+' \cos(2\sqrt{2\pi}\phi_+) ,
\end{split}
\end{equation}
where $g_+'$ is the modified coupling constant 
\begin{equation}
\begin{split}
 g_+' &:= g_+(\ell_1)+\mu_1\gamma_1(\ell_1) \\
 &=\frac{1}{2a} e^{(2-2K_+)\ell_1} \qty[ a_1^2 J_- \Delta+ b_1^2 \mu_1 J_+ e^{-(2K_-)^{-1} \ell_1}].
\end{split}
\end{equation}
Here, we took into account the fact that $g_+$ and $\gamma_1$ terms with the scaling dimensions $2K_+$ and $2K_++(2K_-)^{-1}$, respectively, 
had evolved in the RG flow as exponential functions of $\ell$ with the associated exponents. 

The phase diagram of the sine-Gordon model \eqref{eq:Hp_eff} is well-known \cite{giamarchi2003quantum,gogolin2004bosonization}. 
For $K_+\gtrsim 1$, the cosine potential is irrelevant, and the system is in a critical phase. 
For $K_+\lesssim 1$, the cosine potential is relevant, and the field $\phi_+$ is locked into the minimum of the potential. 
The locking position depends on the sign of $g_+'$, which leads to distinct gapped phases. 
A Gaussian transition between the two gapped phases occurs at $g_+'=0$, i.e., 
\begin{equation}\label{eq:Delta_Gauss}
 \Delta=-\frac{b_1^2 \mu_1 J_+}{a_1^2 J_-} e^{-(2K_-)^{-1} \ell_1} \sim - \frac{J_+}{|J_-|} \qty| \frac{J_-}{J} |^{(4K_--1)^{-1}},
\end{equation}
where we used the fact that $\mu_1$ has the same sign as $\gt_-(0)\propto J_-$. 
According to the celebrated RG argument, BKT transitions from the critical phase 
to the two gapped phases with $\mathrm{sgn}~g_+'=\pm 1$ occur at $y_1=\pm y_2$, where
\begin{equation}\label{eq:y1_y2}
 y_1=2(K_+-1),~y_2=\frac{2\pi \alpha^2 g_+'}{v_+}. 
\end{equation}
In the following analysis, the short-distance cutoff $\alpha$ can be set to the lattice constant $a$. 
The BKT transitions are expected to occur near the XY case $\Delta=0$, where $K_+$ is close to 1. 
In this regime, $y_1$ and $y_2$ can be expressed approximately as
\begin{subequations}\label{eq:y1_y2_approx}
\begin{align}
 y_1&\approx -\frac{4\Delta}{\pi} + \frac{J_+\Delta}{\pi J},\\
 y_2&\approx \frac{\pi}{J} \qty( a_1^2J_-\Delta+b_1^2\mu_1 J_+ e^{-\ell_1/2} ) ,
\end{align}
\end{subequations}
where we set $K_\pm\approx 1$ and $v_+\approx Ja$ in the second equation. 
Therefore, the equations $y_1=\pm y_2$ for the BKT transition lines are solved as
\begin{equation}\label{eq:Delta_BKT}
 \Delta\approx \mp \frac{\pi^2 b_1^2\mu_1J_+}{4J} e^{-\ell_1/2} \sim \mp (\mathrm{sgn}~J_-) \frac{J_+ |J_-|^{1/3}}{J^{4/3}}. 
\end{equation}

In this way, we obtain the schematic phase diagrams in Fig.\ \ref{fig:phases_Delta0}. 
The four gapped featureless phases and two critical phases (XY1 and XY2) compete in a complex manner depending on the signs of $\Jpd$. 
Each gapped featureless phase is characterized by the locking positions of $(2\sqrt{2\pi}\phi_+,\sqrt{2\pi}\theta_-)$; 
see Ref.\ \cite{ogino2021spt} for more details. 
The two critical phases are characterized by the locking of $\theta_-$ in the antisymmetric channel: 
$\sqrt{2\pi}\theta_-=0$ and $\pi$ for XY1 and XY2, respectively. 
In these critical phases, the symmetric channel remains gapless. 
In our field-theoretical prediction, 
the BKT transitions to gapped phases occur at $\Delta<0$ in all the cases in Fig.\ \ref{fig:phases_Delta0}. 
According to the symmetry argument in Ref.\ \cite{PhysRevB.72.014449,Kitazawa_2003}, 
however, the Haldane-XY1 and Haldane*-XY2 transition lines should be located exactly on the $\Delta=0$ line. 
The present approach has a limitation in that the role of such a symmetry in the RG flow is not taken into account. 
The present approach cannot predict the phase structure for $\Jperp\approx 2\Jdiag<0$ and $\Delta\gtrsim 0$ 
(the regimes marked with ``??'' in Fig.\ \ref{fig:phases_Delta0}), either, 
where the $g_+$, $\gt_-$ and $\gamma_1$ terms compete in a nontrivial manner. 
Despite these difficulties, 
the predicted phase diagrams in Fig.\ \ref{fig:phases_Delta0} are remarkably rich, 
and in overall consistency with the phase diagrams obtained by the level spectroscopy 
(Figs.\ \ref{fig:phasediagram_j_jp_model} and \ref{fig:phasediagram_Jxpm0.1}), as we will see in Sec.\ \ref{sec:LevelSpec}

\subsubsection{Case of $\Delta\approx -1$}\label{sec:bos_A_Deltam1}

\begin{figure}
\includegraphics[width=0.45\textwidth]{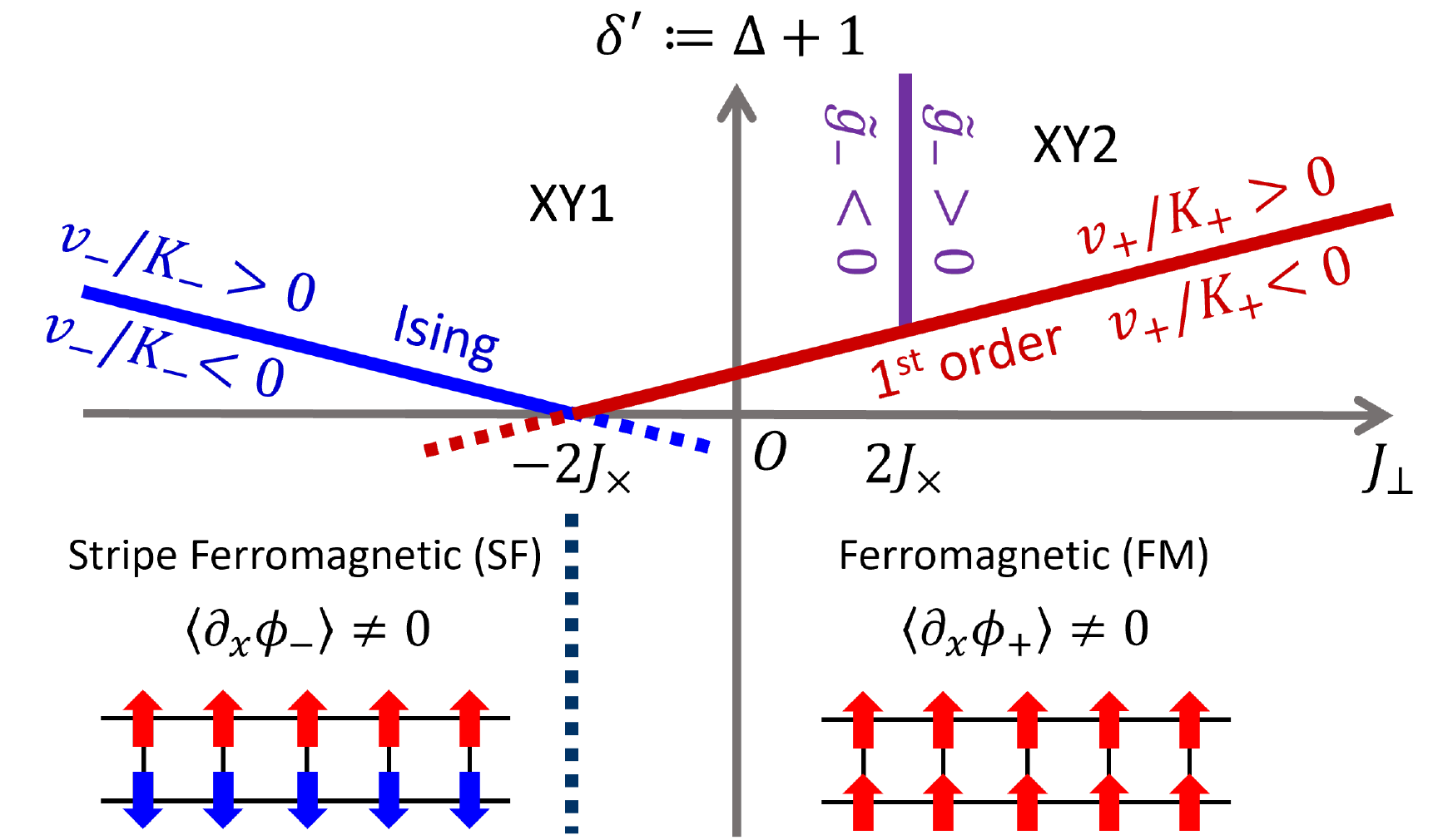} 
\caption{\label{fig:phases_Deltam1}
Schematic phase diagram of the model \eqref{eq:ladfrust_XXZ} for $\Delta=-1+\delta'$ with $|\delta'|\ll 1$, predicted by the effective field theory. 
Here, we fix $\Jdiag$ and vary $\Jperp$ and $\delta'$. 
The transition lines from the critical phases to the FM and SF phases are determined 
from the sign changes of the coefficients $v_\pm/K_\pm$ in Eq.\ \eqref{eq:vpm_o_Kpm} in the Gaussian theory. 
}
\end{figure}

Lastly, we consider the regime of $\Delta=-1+\delta'$ with $|\delta'|\ll 1$, 
where we find different instabilities of the Gaussian theory. 
Specifically, in the effective Hamiltonian \eqref{eq:Heff}, 
the coefficients $v_\pm/K_\pm$ of the terms $(\partial_x \phi_\pm)^2$ have the following expressions in this regime: 
\begin{equation}\label{eq:vpm_o_Kpm}
\begin{split}
 \frac{v_\pm}{K_\pm} 
 = \frac{v}{K}\pm \frac{a}{\pi} J_+ \Delta
 \approx \frac{a}{\pi} \qty( 2J\delta' \mp J_+ ).
\end{split}
\end{equation}
These coefficients become negative for $\delta'<\pm J_+/(2J)$, 
which indicates that the Gaussian theory is unstable. 
If we assume an additional term $(\partial_x\phi_\pm)^4$ with a positive coefficient in the effective Hamiltonian, 
spontaneous magnetic orders with 
\begin{align}
 \langle S_{1,j}^z\pm S_{2,j}^z \rangle 
 = \sqrt{\frac{2}{\pi}} a \langle \partial_x\phi_\pm\rangle\ne 0
\end{align}
are expected to appear along with this instability. 
In this way, ferromagnetic (FM) and stripe ferromagnetic (SF) phases appear 
for $\Jperp>2\Jdiag$ and $\Jperp<2\Jdiag$, respectively, as shown in Fig.\ \ref{fig:phases_Deltam1}. 
Related arguments can be found in Refs.\ \cite{PhysRevB.73.214427,PhysRevA.81.053606}. 

Our estimation of the phase boundaries based on Eq.\ \eqref{eq:vpm_o_Kpm} is naive 
as we neglect the effects of various perturbations in the effective Hamiltonian \eqref{eq:Heff}. 
Indeed, in the numerically obtained phase diagram in Fig.\ \ref{fig:phasediagram_j_jp_model}, 
the transition line to the FM phase is located precisely at $\Delta=-1$, 
and the transition line to the SF phase is bent downward; 
these contrast with Fig.\ \ref{fig:phases_Deltam1}. 
It is also nontrivial to predict the natures of these phase transitions from the field theory. 
However, the transition to the FM phase is clearly of the first order 
as it is associated with the crossing of levels with different total magnetizations $S^z:=\sum_{n,j} S_{n,j}^z$ 
at the ferromagnetic SU(2) point $\Delta=-1$. 
Across this transition, $S^z$ of the ground state changes abruptly from $0$ to $\pm N/2$, 
where $N$ is the number of sites in the system. 
The transition to the SF phase in the XXZ ladder \eqref{eq:ladfrust_XXZ} 
is equivalent to a spontaneous population-imbalance transition in two-component hard-core bosons studied by Takayoshi {\it et al.} \cite{PhysRevA.81.053606}. 
As numerically demonstrated in Ref.\ \cite{PhysRevA.81.053606}, the transition to the population-imbalanced phase
belongs to the Ising universality class in the presence of an intercomponent tunneling 
[i.e., the XY terms in the $\Jperp$ and $\Jdiag$ couplings in Eq.\ \eqref{eq:ladfrust_XXZ}]. 

\section{iDMRG analysis}\label{sec:iDMRG}

In this section, we present numerical results obtained by the iDMRG method \cite{McCulloch_2007, 0804.2509}. 
Our iDMRG calculations are based on Tensor Network Python (TeNPy) \cite{10.21468/SciPostPhysLectNotes.5}. 
We aim to explain how we have obtained the phase diagram for $\Jdiag=-0.5$ in Fig.\ \ref{fig:phasediagram}. 
In particular, we analyze order parameters and topological indices, which can detect symmetry-broken phases and featureless phases, respectively. 

The iDMRG algorithm is based on a periodic matrix product state (MPS) representation of a many-body wave function of the infinite system. 
The precision of this algorithm is controlled by the bond dimension $\chi$. 
In applying this algorithm, we regard the two-leg ladder system as a zigzag chain
arranged in the following order: $\cdots, \Sv_{1,j}, \Sv_{2,j}, \Sv_{1,j+1}, \Sv_{2,j+1}, \cdots$. 
We adopted a four-site unit cell implementation of the iDMRG in calculating the correlation length and the order parameters. 
With this implementation, the variational MPS is expected to converge 
to a symmetry-broken ground state with a four-site (i.e., two-rung) unit cell in the ordered phases (N\'eel, SN, CD, etc.); 
therefore the order parameter can be computed directly. 
The correlation length $\xi$ can be calculated as $\xi(\chi) = - 2/\ln|\epsilon_2(\chi)|$,
where $\epsilon_2(\chi)$ is the second largest absolute eigenvalue of the transfer matrix; 
here, a factor of 2 is multiplied as the transfer matrix is defined over two rungs. 
Meanwhile, we adopted a two-site (i.e., one-rung) unit cell implementation in calculating topological indices in gapped featureless phases 
as a translationally invariant MPS is required for such calculations. 
In most calculations, we exploited the conservation of the total magnetization $\sum_{n,j} S^z_{n,j}=0$ 
to achieve higher efficiency and precision. 
As an exception, we did not use this conservation in analyzing the FM phase (that appears in Fig.\ \ref{fig:phasediagram_j_jp_model} shown later) 
as a finite magnetization spontaneously appears in this phase; see Appendix \ref{app:num_data}. 

\begin{figure}
\includegraphics[width=90mm]{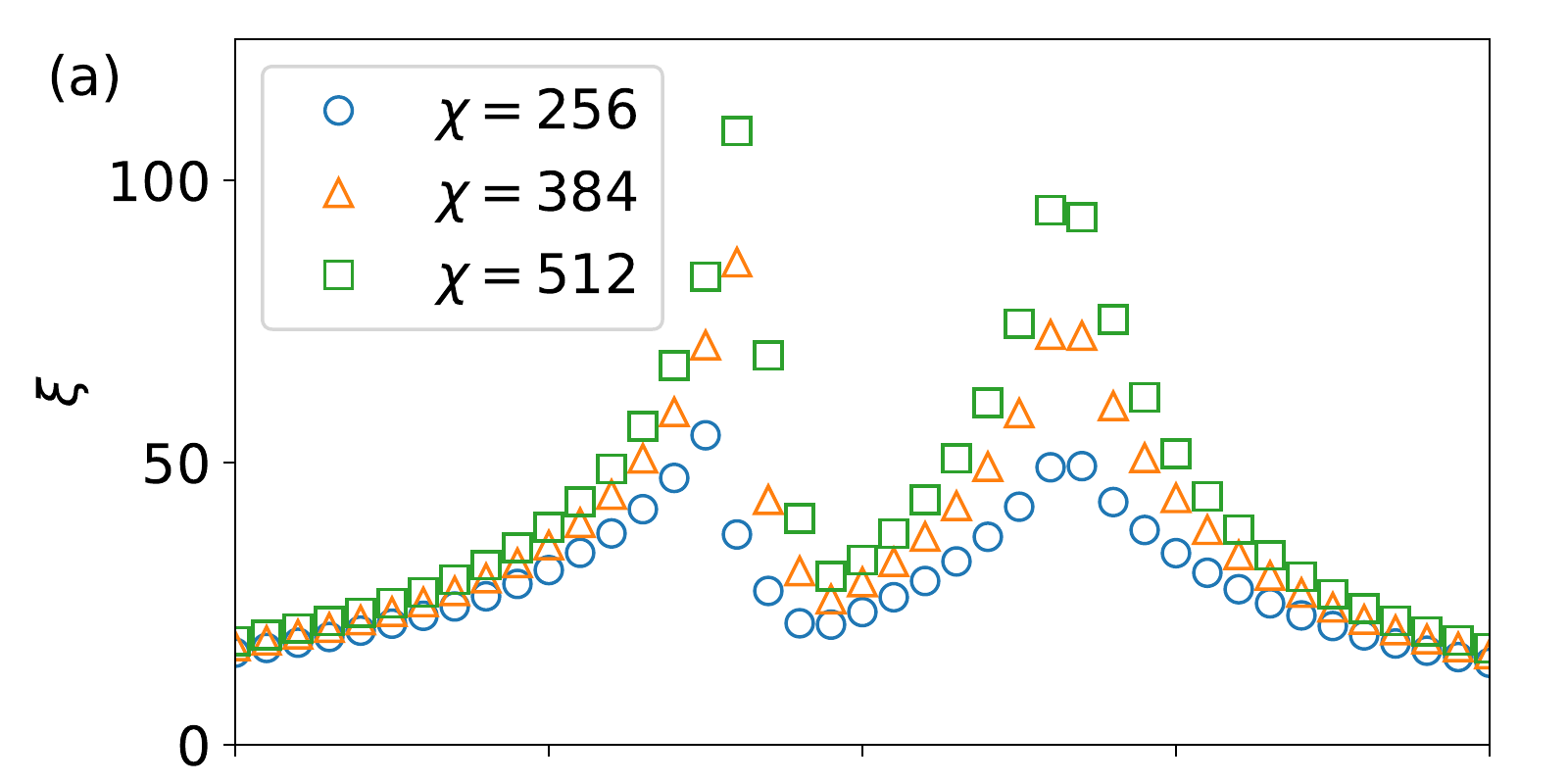}
\includegraphics[width=90mm]{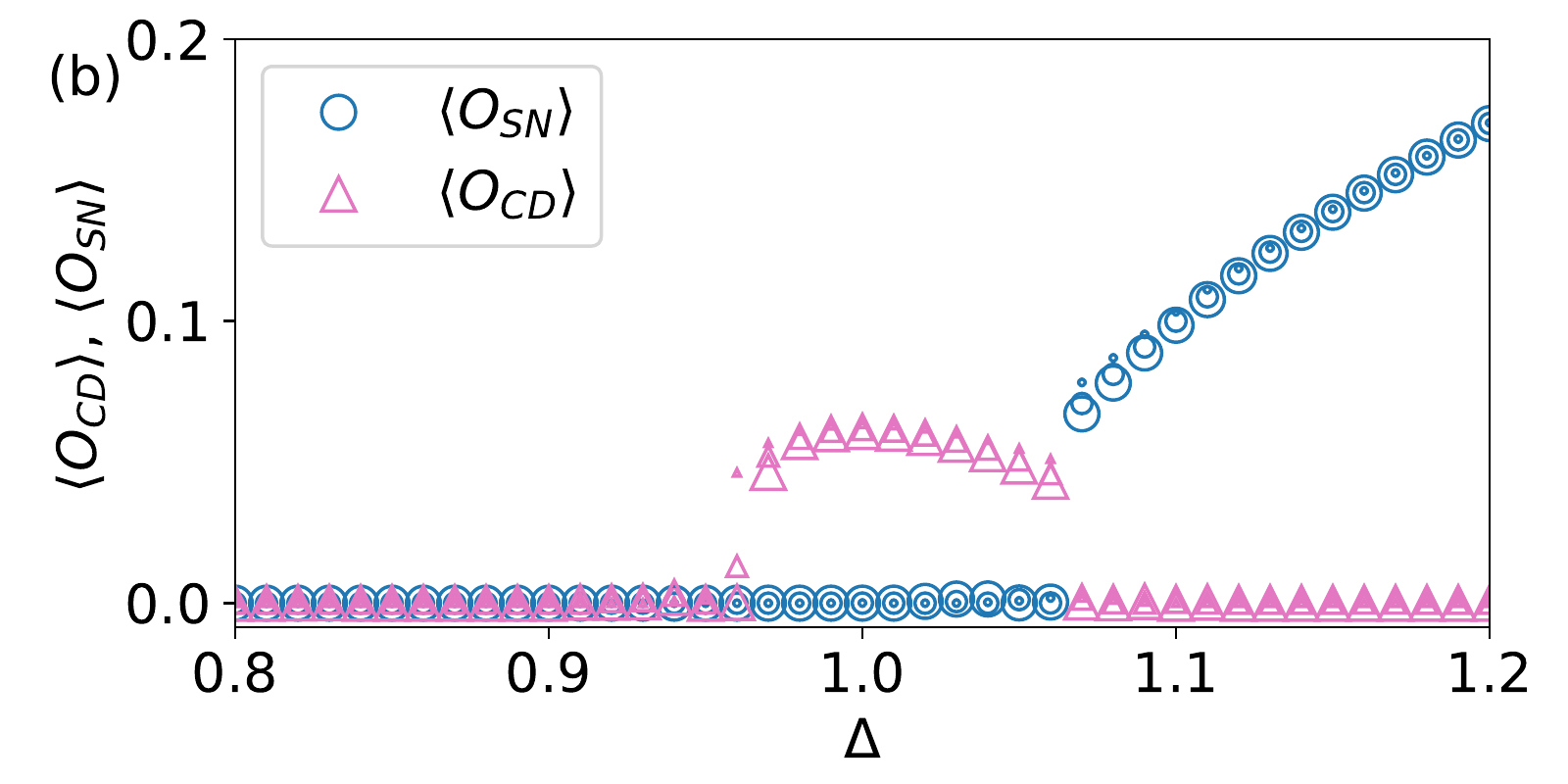}
\caption{
(a) Correlation length $\xi$ and (b) the CD and SN order parameters [Eq.\ \eqref{eq:O_CD_Neel_SN}]
calculated by the iDMRG across the RS*-CD-SN transitions for $J_\perp = -1.1$ and $\Jdiag=-0.5$; see Fig.\ \ref{fig:phasediagram}. 
The correlation length in (a) shows two peaks at $\Delta=0.96$ and $1.06$ that grow with an increase in the bond dimension $\chi$. 
In (b), small, medium, and large symbols are for the bond dimensions $\chi=256$, $384$, and $512$, respectively. 
}
\label{fig:corr_and_order_jpm110}
\end{figure}

\begin{figure}
\includegraphics[width=90mm]{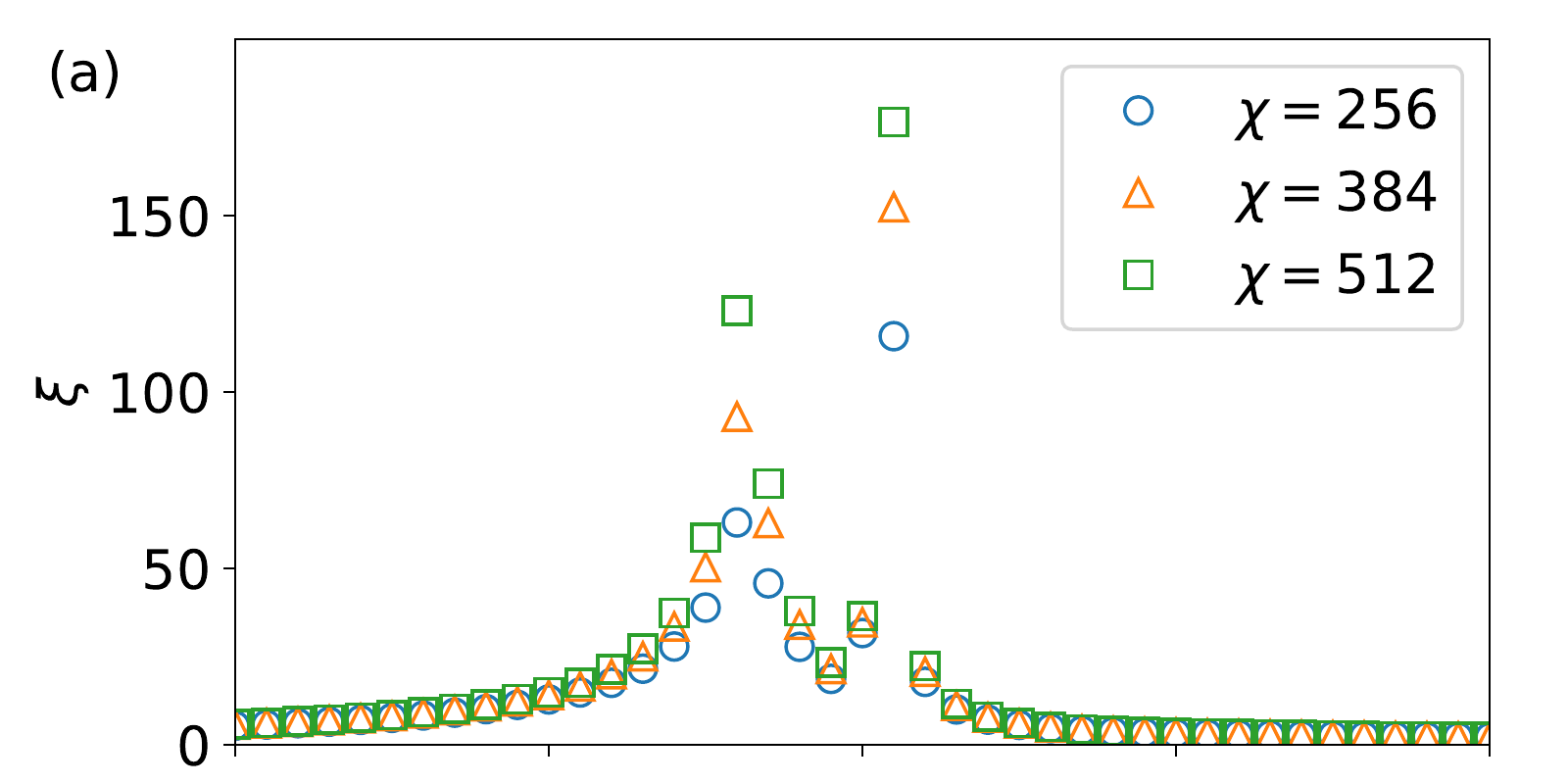}
\includegraphics[width=90mm]{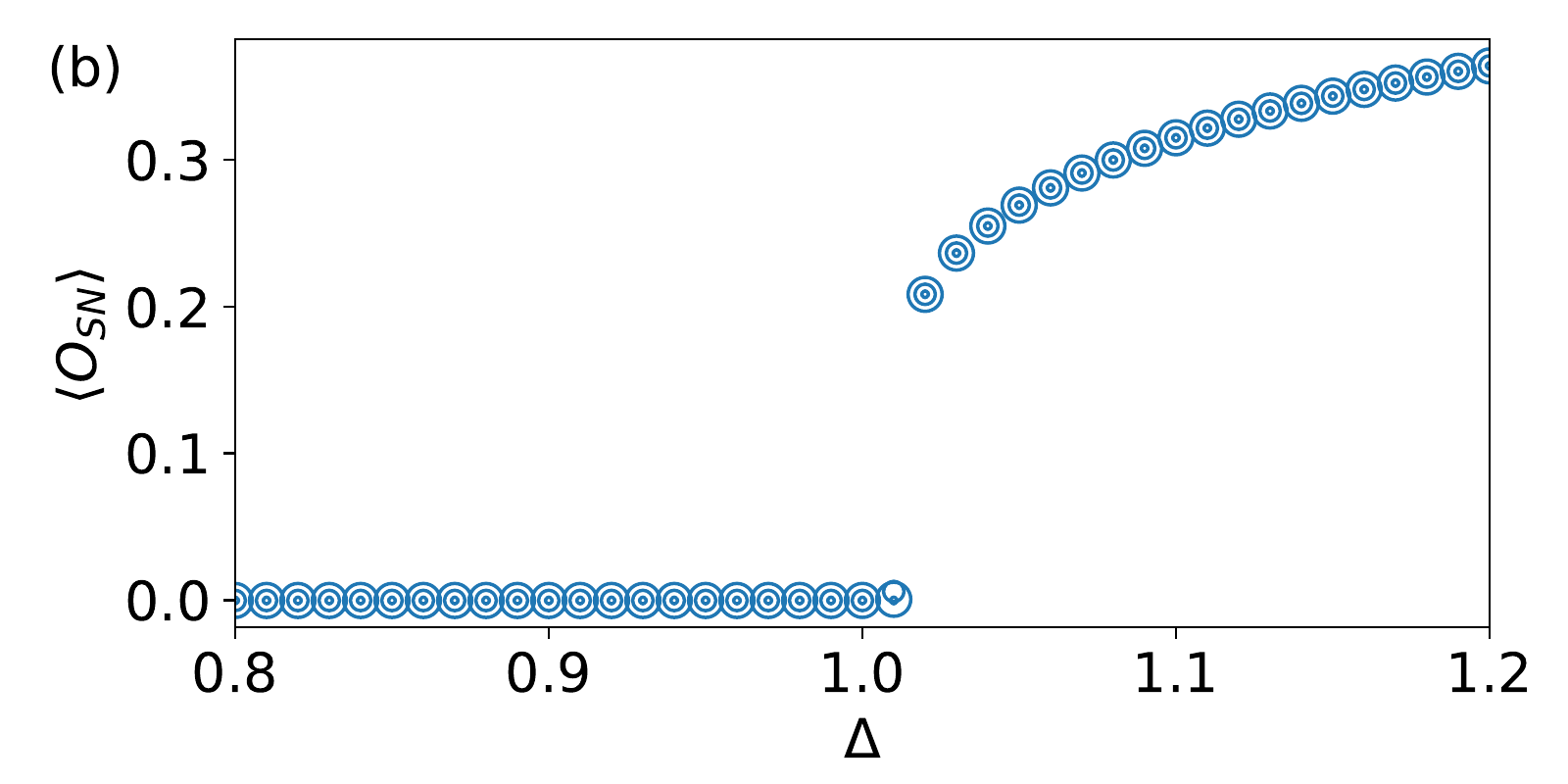}
\caption{
(a) Correlation length $\xi$ and (b) the SN order parameter \eqref{eq:O_SN}
calculated by the iDMRG across the RS*-Haldane-SN transitions for $J_\perp = -2.0$ and $\Jdiag=-0.5$; see Fig.\ \ref{fig:phasediagram}. 
The correlation length in (a) shows two peaks at $\Delta=0.96$ and $1.01$ that grow with an increase in $\chi$. 
In (b), small, medium, and large symbols are for $\chi=256$, $384$, and $512$, respectively. 
}
\label{fig:corr_and_order_jpm200}
\end{figure}

\subsection{Correlation length and order parameters}\label{sec:Orders}

Symmetry-broken phases can be identified by calculating the associated order parameters. 
The N\'eel, SN, and CD order parameters are given by 
\begin{subequations}\label{eq:O_CD_Neel_SN}
\begin{align}
\expval{\mathcal{O}_{\text{N\'eel}}(j)} &= \frac14 \expval{S^z_{1,j} - S^z_{2,j} - S^z_{1,j+1} + S^z_{2,j+1}}, \label{eq:O_Neel}\\
\expval{\mathcal{O}_{\text{SN}}(j)} &= \frac14 \expval{S^z_{1,j} + S^z_{2,j} - S^z_{1,j+1} - S^z_{2,j+1}}, \label{eq:O_SN}\\
\expval{\mathcal{O}_{\text{CD}}(j)} &= \frac14 
\langle \bm{S}_{1,j-1}\cdot\bm{S}_{1,j} + \bm{S}_{2,j-1}\cdot\bm{S}_{2,j} \notag\\
&~~~~~~-\bm{S}_{1,j}\cdot\bm{S}_{1,j+1} - \bm{S}_{2,j}\cdot\bm{S}_{2,j+1} \rangle. \label{eq:O_CD} 
\end{align}
\end{subequations}

\begin{figure}
\includegraphics[width=0.45\textwidth]{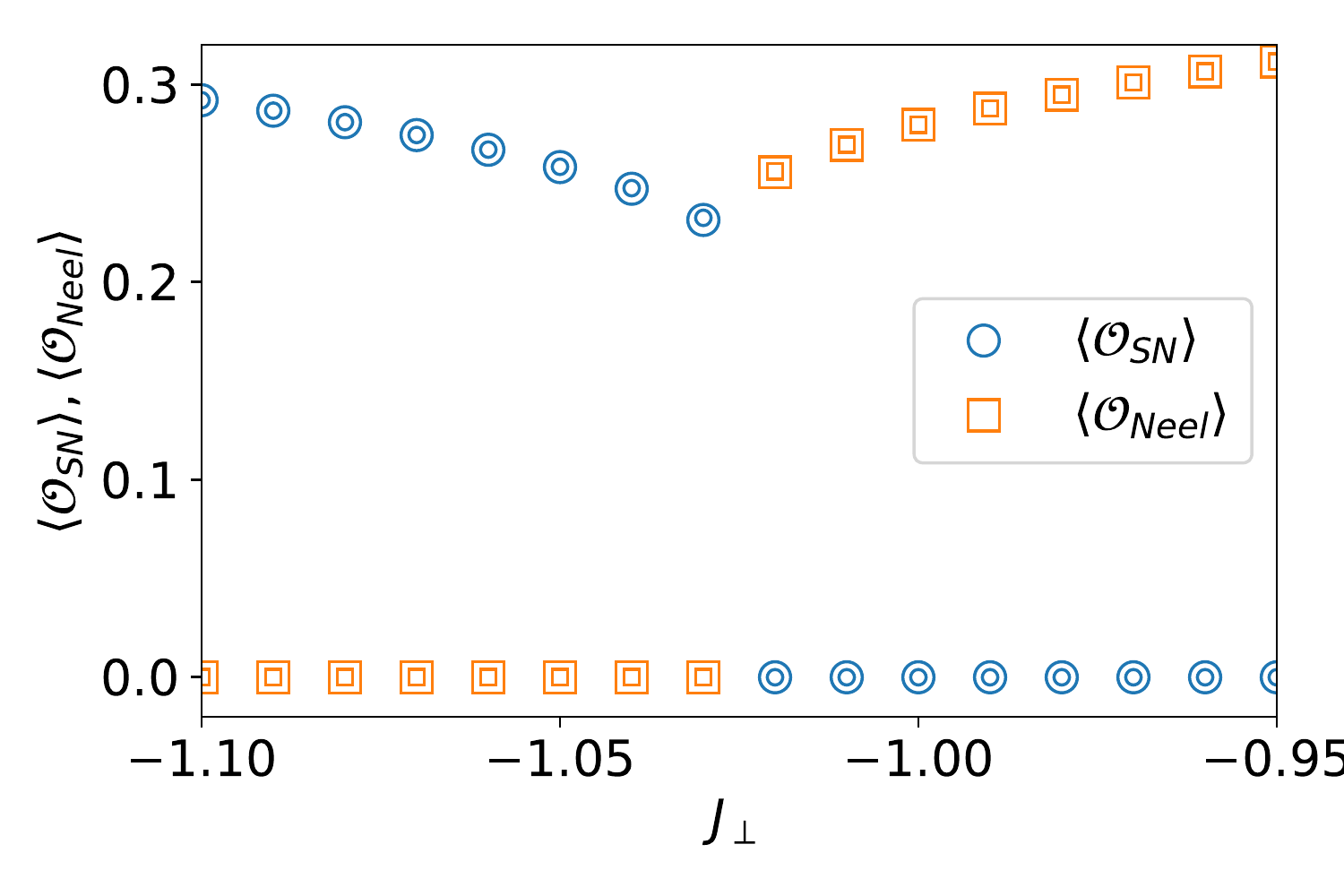}
\caption{
SN and N\'eel order parameters [Eq.\ \eqref{eq:O_CD_Neel_SN}] calculated by the iDMRG 
across the SN-N\'eel transition for $\Delta=1.5$ and $\Jdiag=-0.5$; see Fig.\ \ref{fig:phasediagram}. 
Small and large symbols are for $\chi=128$ and $\chi=256$, respectively. 
Sudden changes in the calculated order parameters between $\Jperp=-1.03$ and $-1.02$ indicate a first-order phase transition.  
}
\label{fig:order_SN_Neel}
\end{figure}

Figure \ref{fig:corr_and_order_jpm110} shows the correlation length $\xi$ and 
the CD and SN order parameters across the RS*-CD-SN transitions. 
The correlation length in Fig.\ \ref{fig:corr_and_order_jpm110}(a) shows two peaks that grow with an increase in the bond dimension $\chi$. 
It is therefore likely that second-order transitions with divergence in the correlation length occur at these points. 
The peak positions $\Delta=0.96$ and $1.06$ are consistent with the onset of the CD and SN order parameters shown in Fig.\ \ref{fig:corr_and_order_jpm110}(b). 
We therefore use the peak positions for largest $\chi$ as the estimation of the transition points. 

Figure \ref{fig:corr_and_order_jpm200} shows the correlation length $\xi$ and the SN order parameter 
across the RS*-Haldane-SN transitions. 
Again, the correlation length in Fig.\ \ref{fig:corr_and_order_jpm200}(a) shows 
two peaks with growing heights, which indicate second-order transitions. 
The peak at $\Delta=1.01$ is consistent with the onset of the SN order parameter in Fig.\ \ref{fig:corr_and_order_jpm200}(b). 
Meanwhile, the RS* and Haldane phases are featureless, and cannot be distinguished by a local order parameter. 
These phases are distinguished by topological indices as explained in Sec.\ \ref{sec:topo_indices}. 

According to the effective field theory analysis in Sec.\ \ref{sec:bos_A_Delta1}, 
the CD, SN, Haldane, and RS* phases are separated by Gaussian and Ising transitions, as shown in Fig.\ \ref{fig:phases_Delta1}. 
A numerical verification of these universality classes is beyond the scope of our present study. 
In both Figs.\ \ref{fig:corr_and_order_jpm110} and \ref{fig:corr_and_order_jpm200}, 
the two critical points are relatively close to each other. 
Therefore, in examining the universality class of one of these critical points by the iDMRG with finite $\chi$, 
there is an unavoidable influence from the other critical point. 
However, we note that the universality classes have been examined in detail in an XXZ ladder with an explicit dimer attraction term \cite{ogino2021spt}, 
in which crossing of Gaussian and Ising transition lines similar to Fig.\ \ref{fig:phases_Delta1} has been confirmed. 

Figure \ref{fig:order_SN_Neel} shows the SN and N\'eel order parameters across the SN-N\'eel transition in the easy-axis regime. 
Sudden changes in the calculated order parameters between $\Jperp=-1.03$ and $-1.02$ indicate a first-order phase transition. 
As discussed in Sec.\ \ref{sec:bos_A_easyaxis}, a first-order transition between the SN and N\'eel phases 
is consistent with the simplified effective Hamiltonian in which $\tilde{g}_-$ and $\gamma_1$ are neglected. 

In the phase diagram for $\Jdiag=-0.5$ in Fig.\ \ref{fig:phasediagram}, the N\'eel, SN, and CD phases are identified 
by examining the robustness of the associate order parameters with an increase in the bond dimension $\chi$.

\subsection{Topological indices}\label{sec:topo_indices}

The four featureless phases (RS, RS*, Haldane, and Haldane*) are all distinct 
in the presence of the $D_2\times\sigma$ and translational symmetries \cite{PhysRevB.86.195122, Fuji15, ogino2021spt}, 
where $D_2=\mathbb{Z}_2\times\mathbb{Z}_2$ is the discrete spin rotational symmetry and $\sigma$ is the leg-interchange symmetry. 
Here, we analyze topological indices \cite{PhysRevB.81.064439, PhysRevB.86.125441, PhysRevB.83.035107, PhysRevB.84.235128} 
associated with these symmetries, which allow us to identify the four featureless phases unambiguously. 

Details on how to define and calculate topological indices for the present ladder setup has been described in Ref.\ \cite{ogino2021spt}. 
Here, we summarize the basic idea and the definitions. 
Assuming the translational invariance, we represent the ground state of the infinite system in the form of a canonical MPS as 
\begin{align}\label{eq:uMPS}
\ket{\Psi} = \sum_{\dots,l,m,n,\dots} 
[ \dots \Lambda \Gamma_{l} \Lambda \Gamma_{m} \Lambda \Gamma_{n} \dots] \ket{\dots l,m,n, \dots}. 
\end{align}
Here, $\Gamma_m$ is a $\chi$-by-$\chi$ matrix with $m$ being the spin state on a ``site'',
and $\Lambda=\mathrm{diag}(\lambda_1,\dots,\lambda_\chi)$ is a diagonal matrix comprised of Schmidt values. 
In our application to the spin-$\frac12$ ladder, a ``site'' corresponds to a rung, and $m$ runs over the four spin states on a rung
\footnote{
In our iDMRG calculations, we employed a periodic MPS representation along the zigzag chain as noted at the beginning of Sec.\ \ref{sec:iDMRG}. 
In this representation, a tensor is defined for each spin. 
We combined tensors at each rung into a larger tensor $\Gamma_m$ to obtain the representation in Eq.\ \eqref{eq:uMPS}. 
}. 
Suppose that $\ket{\Psi}$ is invariant under an on-site unitary transformation, 
which is represented in the spin basis as a unitary matrix $\Sigma_{mm'}$ acting on every ``site''. 
Then the $\Gamma_m$ matrices can be shown to satisfy \cite{PhysRevLett.100.167202} 
\begin{align}\label{eq:defU}
\sum_{m'} \Sigma_{mm'} \Gamma_{m'} = e^{i\theta_\Sigma} U_\Sigma^\dagger \Gamma_m U_\Sigma,
\end{align}
where $e^{i\theta_\Sigma}$ is a phase factor and $U_\Sigma$ is a $\chi$-by-$\chi$ unitary matrix which commutes with $\Lambda$. 
The matrix $U_\Sigma$ can be viewed as a symmetry transformation acting on fictitious edge states in the entanglement Hamiltonian. 

\begin{figure}
\includegraphics[width=90mm]{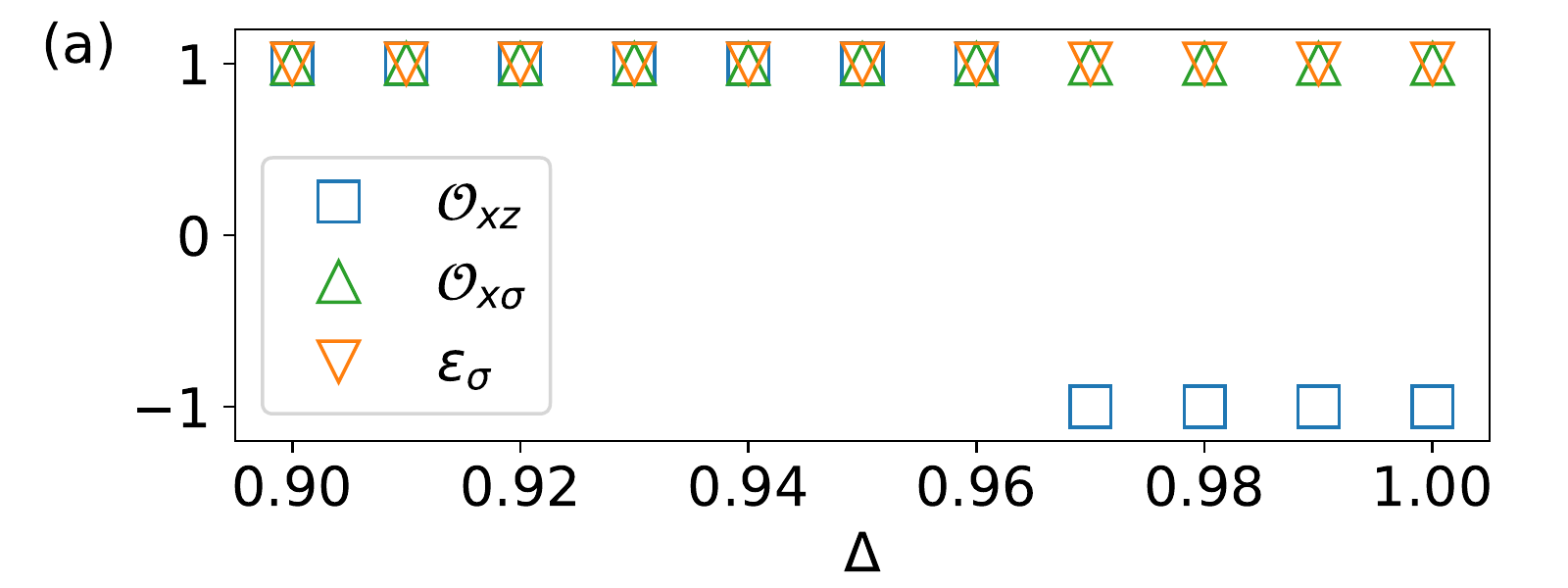}
\includegraphics[width=90mm]{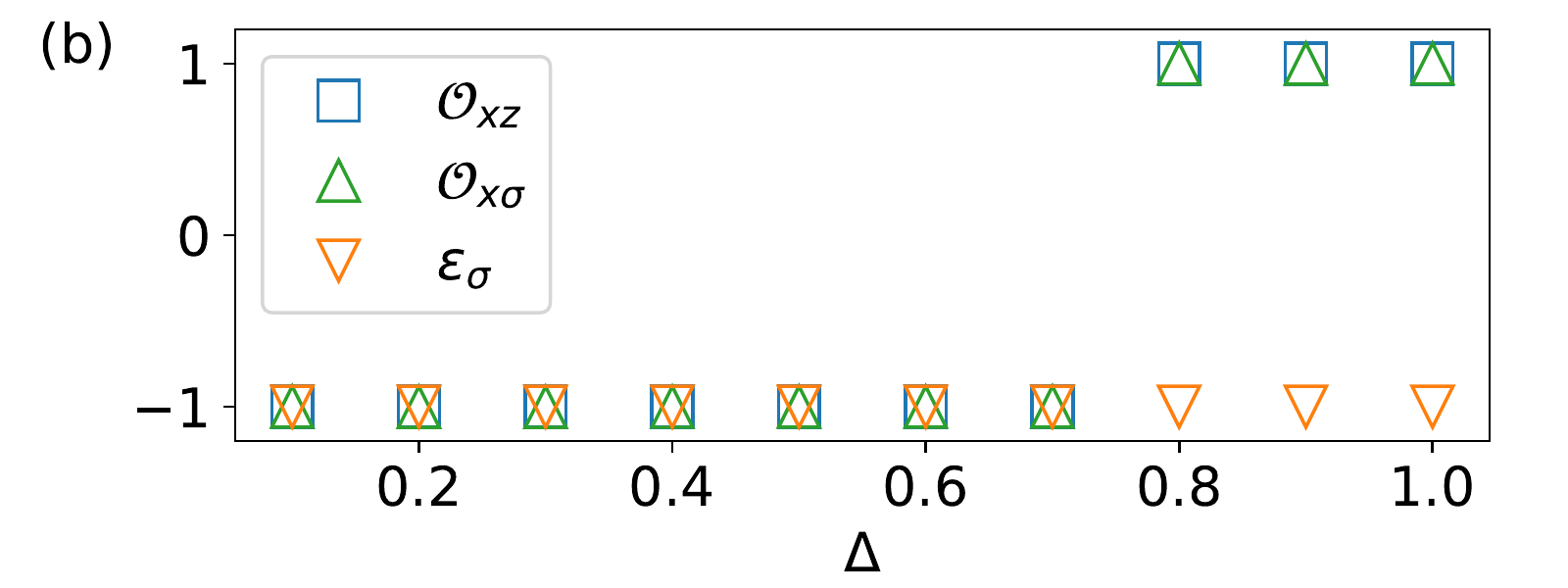}
\caption{
Topological indices calculated by the iDMRG with $\chi=256$ 
across (a) the RS*-Haldane transition for $\Jperp=-2.0$ and $\Jdiag=-0.5$
and (b) the Haldane*-RS transition for $\Jperp=\Jdiag=-0.5$; see Fig.\ \ref{fig:phasediagram}. 
The indices $\mathcal{O}_{xz}$ and $\mathcal{O}_{x\sigma}$ are the traced commutators in Eq.\ \eqref{eq:Oxz_Oxsig} 
while $\epsilon_1^\sigma:=e^{i\theta_{\Sigma_\sigma}}$ is the phase factor $e^{i\theta_\Sigma}$ in Eq.\ \eqref{eq:defU} for the leg interchange $\sigma$. 
}
\label{fig:topo}
\end{figure}

For a spin ladder with the $D_2\times\sigma$ symmetry, we consider the following on-site transformations: 
the spin rotations $\Sigma^x := \exp[i\pi (S^x_1 + S^x_2)]$ and $\Sigma^z := \exp[i\pi (S^z_1 + S^z_2)]$ 
and the leg interchange $\Sigma^\sigma:=\sum_{\alpha,\beta=\uparrow,\downarrow}\ket{\alpha\beta}\bra{\beta\alpha}$. 
For these transformations, we can introduce the unitary matrices 
$U_x:=U_{\Sigma^x}$, $U_z:=U_{\Sigma^z}$, and $U_\sigma:=U_{\Sigma^\sigma}$ as in Eq.\ \eqref{eq:defU}. 
While $\Sigma^x$, $\Sigma^z$, and $\Sigma^\sigma$ commute with one another, 
the commutation relations among $U_x$, $U_z$, and $U_\sigma$ may involve nontrivial phase factors. 
Such phase factors can be used to distinguish different featureless phases. 
They can conveniently be detected by calculating traced commutators \cite{PhysRevB.86.125441} 
\begin{subequations}\label{eq:Oxz_Oxsig}
\begin{align}
\mathcal{O}_{xz} &:= \frac{1}{\chi} \tr( U_x U_z U_x^\dagger U_z^\dagger ),\\
\mathcal{O}_{x\sigma} &:= \frac{1}{\chi} \tr( U_x U_\sigma U_x^\dagger U_\sigma^\dagger ).
\end{align}
\end{subequations}
For example, we have $\mathcal{O}_{xz}=-1$ from the nontrivial relation $U_x U_z = - U_z U_x$ in the Haldane and Haldane* phases. 
We have $\mathcal{O}_{x\sigma}=-1$ from the nontrivial relation $U_x U_\sigma = - U_\sigma U_x$ in the Haldane* phase. 
We also examine the phase factor $\epsilon_1^\sigma:=e^{i\theta_{\Sigma_\sigma}}$ in Eq.\ \eqref{eq:defU} for the leg interchange $\sigma$. 
This index can distinguish the RS and RS* phases (with $\epsilon_1^\sigma=-1$ and 1, respectively) in the presence of the translational symmetry. 
In this way, we can distinguish all the four featureless phases by using the three indices. 

Figure \ref{fig:topo} shows the topological indices $\mathcal{O}_{xz}$, $\mathcal{O}_{x\sigma}$, and $\epsilon_1^\sigma$ 
calculated by the iDMRG across (a) the RS*-Haldane transition and (b) the Haldane*-RS transition. 
In Fig.\ \ref{fig:topo}(a), we find a sudden change in $\mathcal{O}_{xz}$ between $\Delta=0.96$ and $0.97$. 
This is consistent with the peak of the correlation length at $\Delta=0.96$ in Fig.\ \ref{fig:corr_and_order_jpm110}(a).
In Fig.\ \ref{fig:topo}(b), we find sudden changes in $\mathcal{O}_{xz}$ and $\mathcal{O}_{x\sigma}$ between $\Delta=0.7$ and $0.8$. 
For this transition, we have also performed a level spectroscopy analysis, 
and obtained the estimate $\Delta=0.710$ of the transition point in the thermodynamic limit; 
see Fig.\ \ref{fig:levelspec_Jrm050_Jcm050} in Appendix \ref{app:num_data}. 
We can thus confirm a consistency between the independent analyses. 

In the phase diagram for $\Jdiag=-0.5$ in Fig.\ \ref{fig:phasediagram}, the four gapped featureless phases 
are identified by examining the topological indices as in Fig. \ref{fig:topo}. 
It is worth emphasizing that our iDMRG results on the correlation length, the order parameters, and the topological indices 
consistently support the rich phase structure around the isotropic case $\Delta=1$ in Fig.\ \ref{fig:phases_Delta1}. 

\section{Level spectroscopy}\label{sec:LevelSpec}

\begin{figure}
\includegraphics[width=0.45\textwidth]{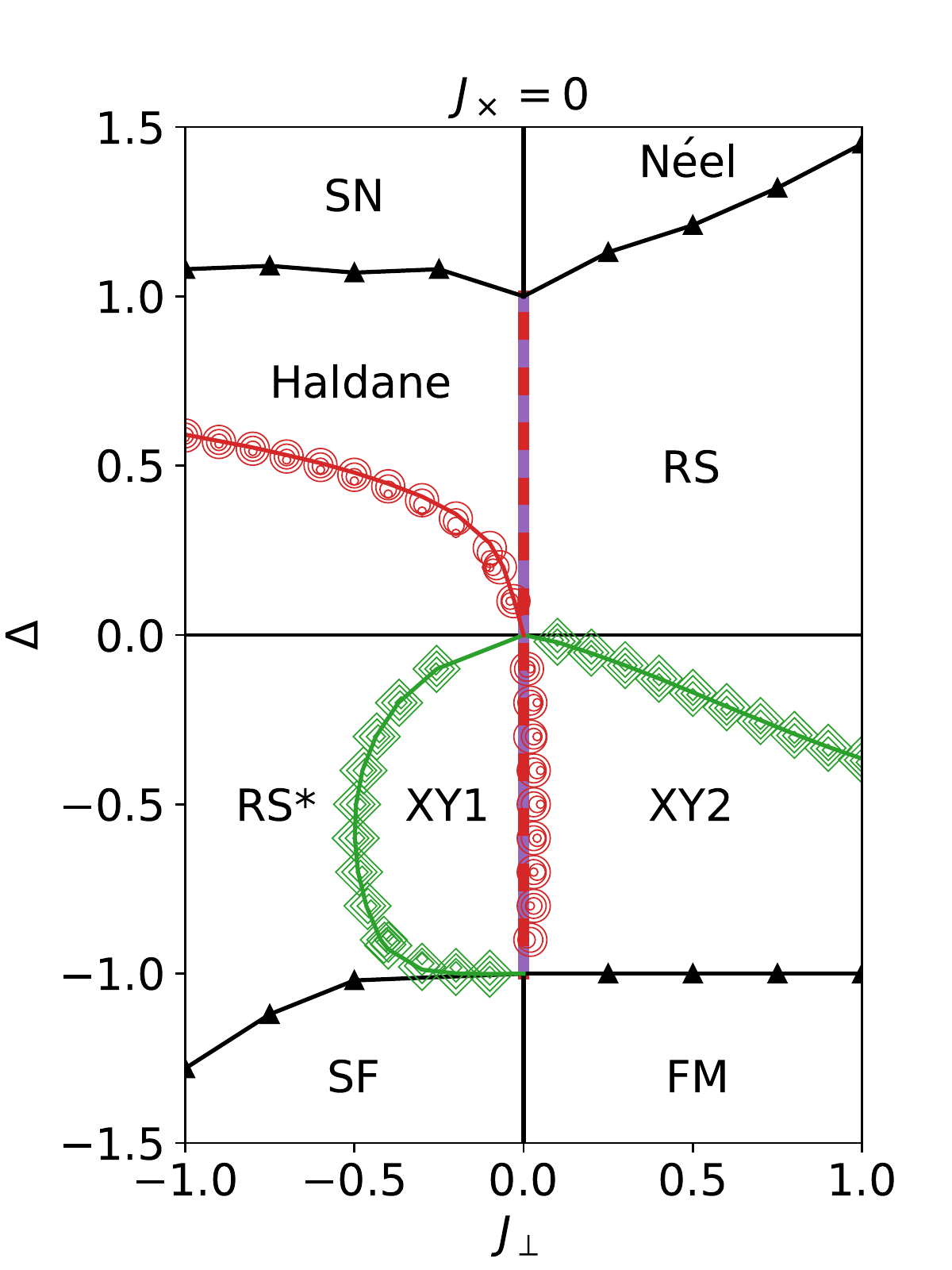}
\caption{\label{fig:phasediagram_j_jp_model}
Phase diagram of the simple XXZ ladder model, i.e., Eq.\ \eqref{eq:ladfrust_XXZ} with $\Jdiag=0$, obtained numerically. 
The phase boundaries in the easy-plane regime $-1<\Delta<1$ are determined by the level spectroscopy analysis. 
Red circles and green rhombuses indicate level crossing points corresponding to Gaussian and BKT lines. 
Here, four different symbol sizes (from small to large) correspond to $L=6$, $8$, $10$, and $12$. 
The data extrapolated to thermodynamic limit are connected by red and green lines. 
Black triangles indicate transition points determined by the iDMRG analysis of the order parameters. 
Specifically, the boundaries of the Neel and SN phases are determined in a similar manner as in Sec.\ \ref{sec:Orders}; 
the boundary of the FM phase is determined in Appendix \ref{app:num_data}. 
The vertical line with alternating red and purple colors at $\Jperp=0$ 
indicate sign changes of $g_+$ and $\gt_-$ in the effective Hamiltonian \eqref{eq:Heff}; 
we have also confirmed the corresponding level crossings precisely on this line in the level spectroscopy analysis (not shown). 
See Fig.\ \ref{fig:phases_Delta0}(a) for a phase diagram based on the effective field theory,  
and Ref.\ \cite{Li_2017} for a previous tensor network study. 
We note that red circles for $\Delta<0$ do not correspond to a phase transition 
but are shown for reference as they can be compared with the $y_2=0$ line in Fig.\ \ref{fig:phases_Delta0}(a). 
}
\end{figure}

\begin{figure}
\includegraphics[width=0.45\textwidth]{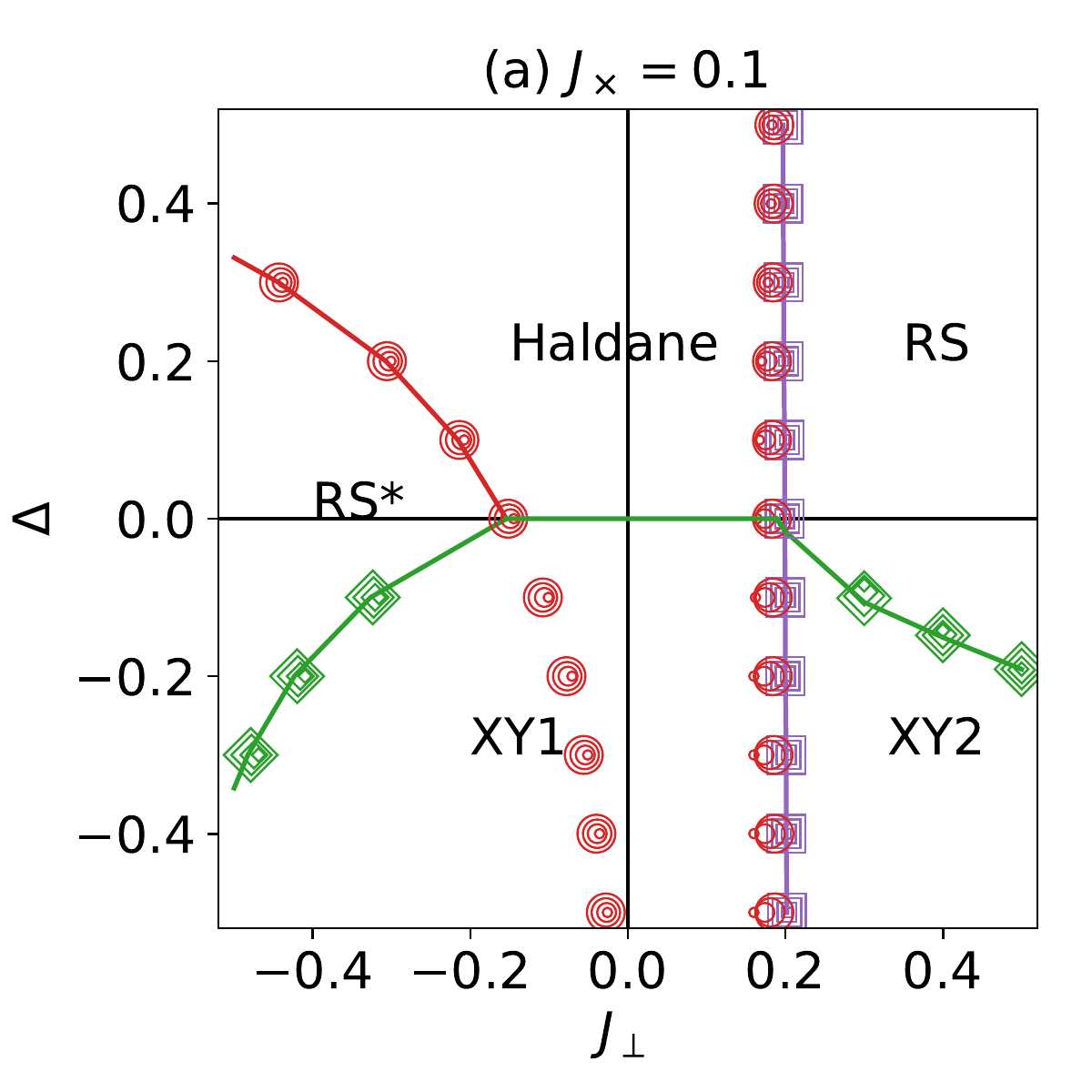}
\includegraphics[width=0.45\textwidth]{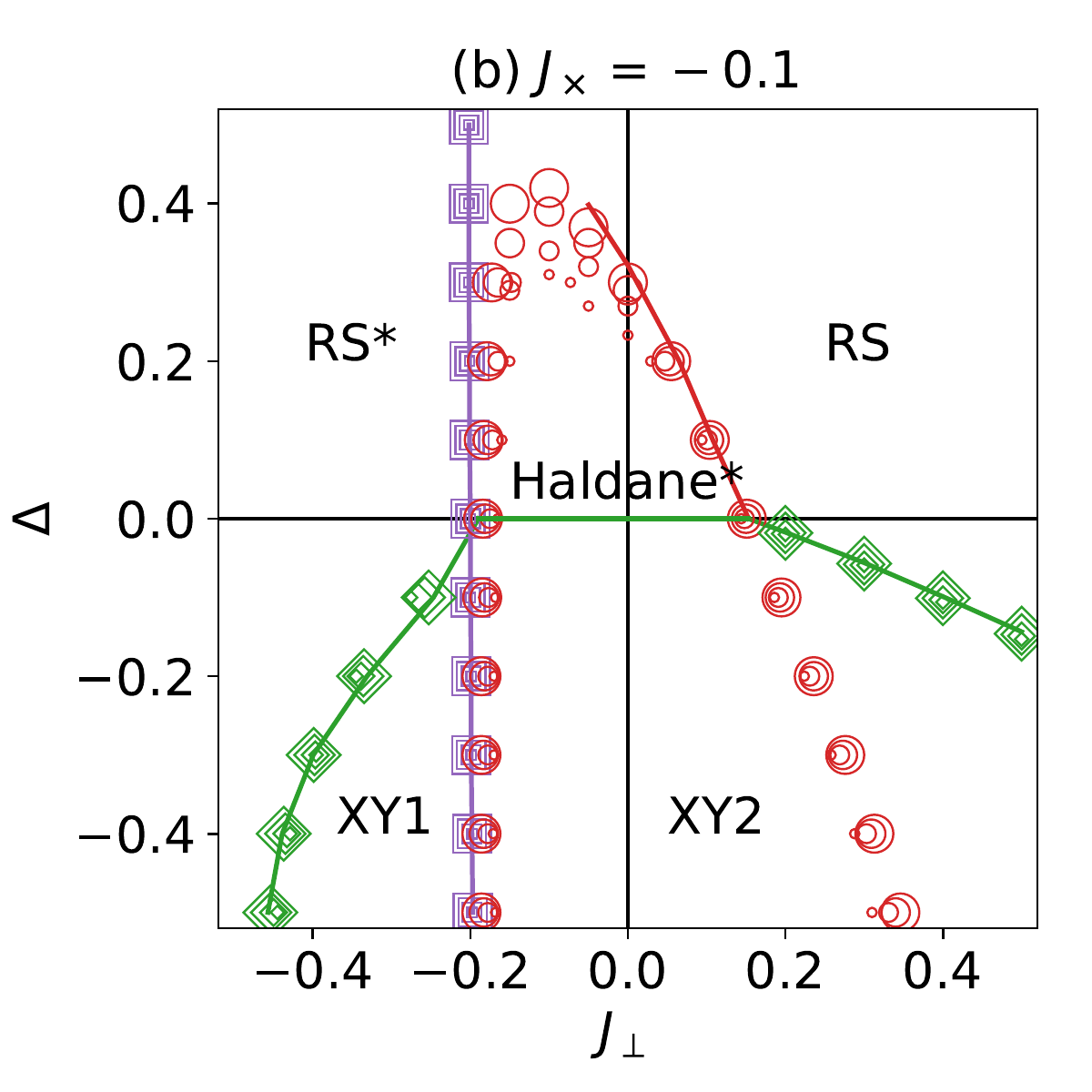}
\caption{\label{fig:phasediagram_Jxpm0.1}
Phase diagrams of the model \eqref{eq:ladfrust_XXZ} with (a) $\Jdiag=0.1$ and (b) $\Jdiag=-0.1$, obtained by the level spectroscopy analysis. 
Green rhombuses indicate level crossing points corresponding to BKT transition lines. 
Red circles indicate level crossing of $\Delta\Erm{H}$ and $\Delta\Erm{R}$, as shown in Figs.\ \ref{fig:levelspec_D020}(a) and \ref{fig:levelspec_Dm020}(a). 
For $\Delta>0$ and $\Jperp$ away from $2\Jdiag$, this level crossing corresponds to a Haldane-RS* or Haldane*-RS Gaussian transition. 
Purple squares indicate level crossing of $\Erm{XY1}$ and $\Erm{XY2}$, as shown in Fig.\ \ref{fig:levelspec_Dm020}(b). 
Here, four different symbol sizes (from small to large) correspond to $L\in\{6,8,10,12\}$ for green and red 
and $L\in\{5,7,9,11\}$ for purple. 
The data extrapolated to thermodynamic limit are connected by colored lines. 
Based on the symmetry argument in Ref.\ \cite{PhysRevB.72.014449,Kitazawa_2003}, 
the Haldane-XY1 and Haldane*-XY2 transition lines are shown by horizontal green lines at $\Delta=0$. 
See Figs.\ \ref{fig:phases_Delta0}(b,c) for phase diagrams based on the effective field theory. 
We note that the red circles for $\Delta<0$ do not correspond to a phase transition 
but can be compared with the $g_+=0$ and $y_2=0$ lines in Figs.\ \ref{fig:phases_Delta0}(b,c). 
}
\end{figure}

In this section, we present a level spectroscopy analysis, 
which combines effective field theory with numerical exact diagonalization. 
Our exact diagonalization calculations are based on the Python package QuSpin \cite{SciPostPhys.2.1.003,10.21468/SciPostPhys.7.2.020}. 
We aim to numerically demonstrate the rich phase structures in the easy-plane regime predicted by the bosonization (Fig.\ \ref{fig:phases_Delta0}), 
where the four featureless phases and the two critical phases compete in a complex manner. 
In the following analysis, we set $\Jdiag$ to relatively small values in magnitudes 
so that the numerical results can be compared directly with the bosonization results for $|\Jpd|\ll J$. 
The phase diagrams obtained numerically for $\Jdiag=0$ and $\pm 0.1$ are shown in Figs.\ \ref{fig:phasediagram_j_jp_model} and \ref{fig:phasediagram_Jxpm0.1}. 
Below we explain how the phase boundaries in the easy-plane regime $-1<\Delta<1$ in these diagrams are obtained by the level spectroscopy. 


\begin{figure}
\includegraphics[width=0.45\textwidth]{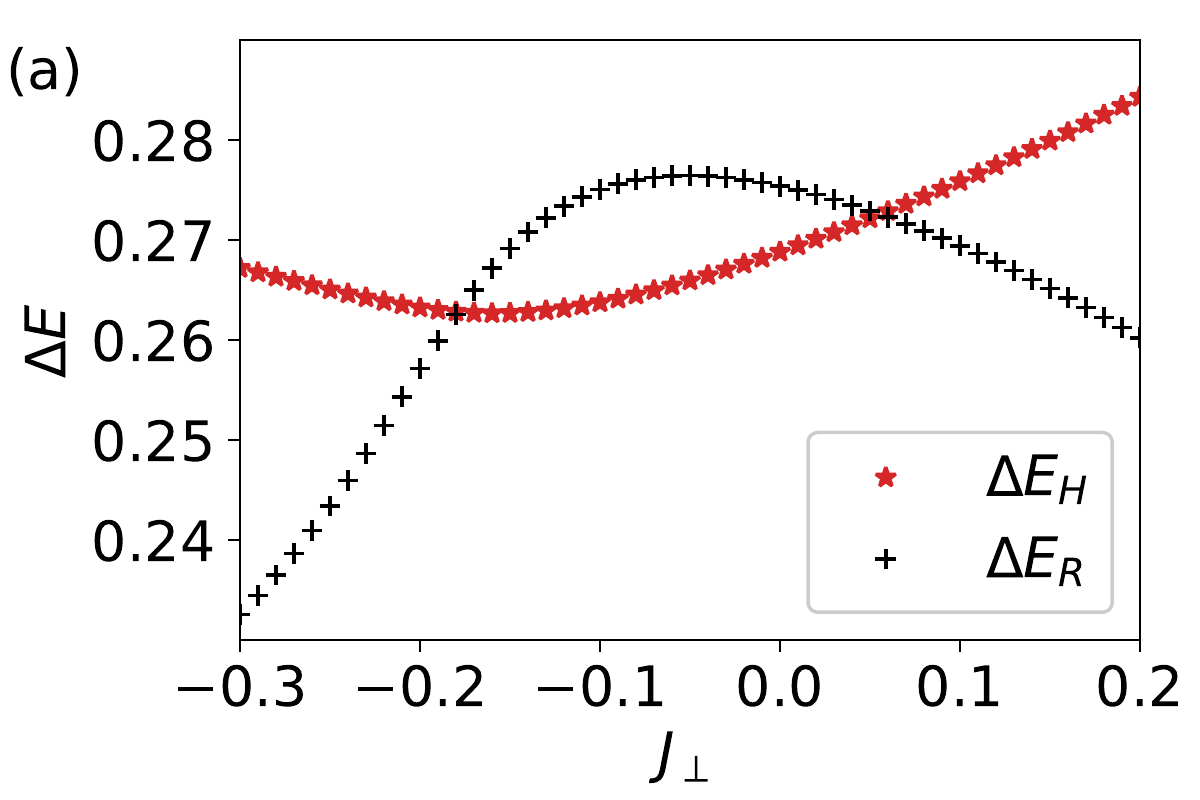}
\includegraphics[width=0.45\textwidth]{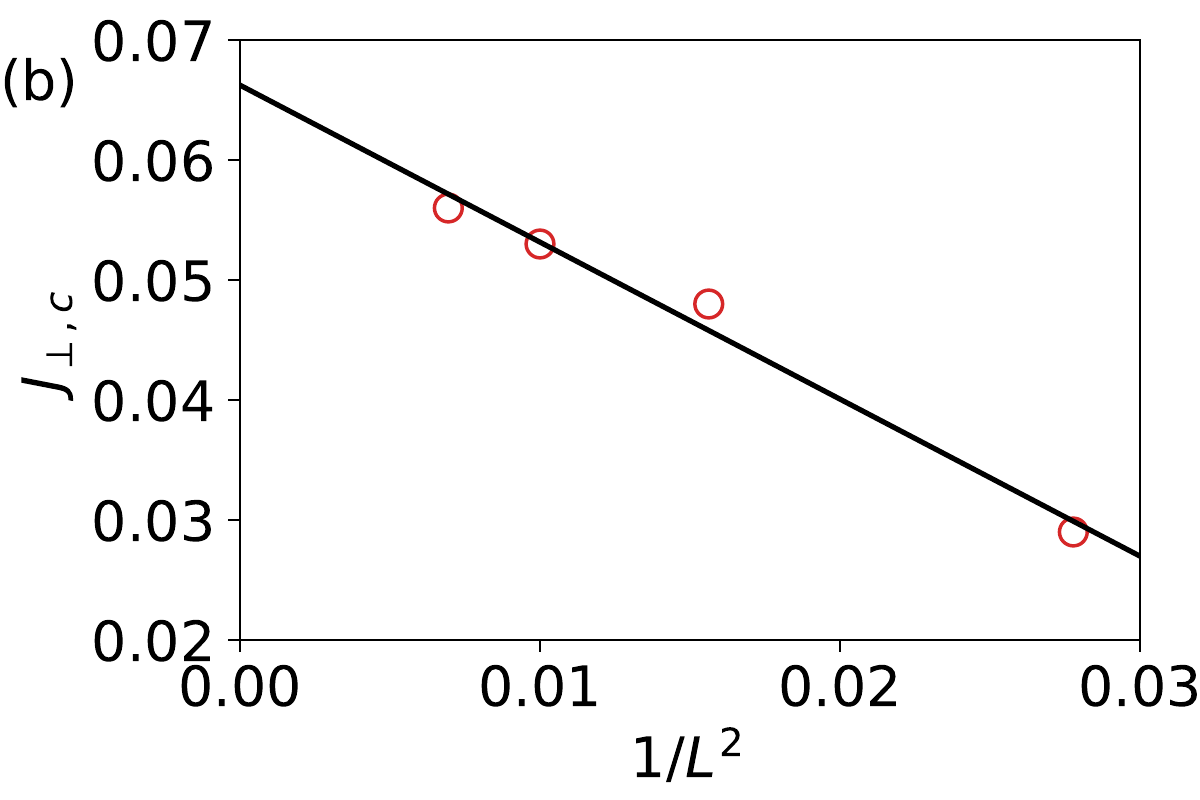}
\caption{
(a) Excitation energies $\Delta\Erm{H}$ and $\Delta\Erm{R}$ in Eq.\ \eqref{eq:E_XYHR} 
calculated by exact diagonalization ($L=12$) 
for $\Delta=0.2$ and $\Jdiag=-0.1$; see Fig.\ \ref{fig:phasediagram_Jxpm0.1}(b). 
The crossing point at $\Jperp=J_{\perp,c}^\text{H*-R}=0.056$ gives a finite-size estimate of the Haldane*-RS transition. 
Meanwhile, the crossing point at $\Jperp =-0.179$ gives a rough estimate of the left boundary of the Haldane* phase; see the text for details.
(b) Crossing point $J_{\perp,c}^\text{H*-R}(L)$ versus $1/L^2$. 
An extrapolation using a linear function of $1/L^2$ yields 
the estimate $J_{\perp,c}^\text{H*-R}(L\to\infty)=0.066$ of the transition point in the thermodynamic limit. 
}
\label{fig:levelspec_D020}
\end{figure}

\begin{figure}
\includegraphics[width=0.45\textwidth]{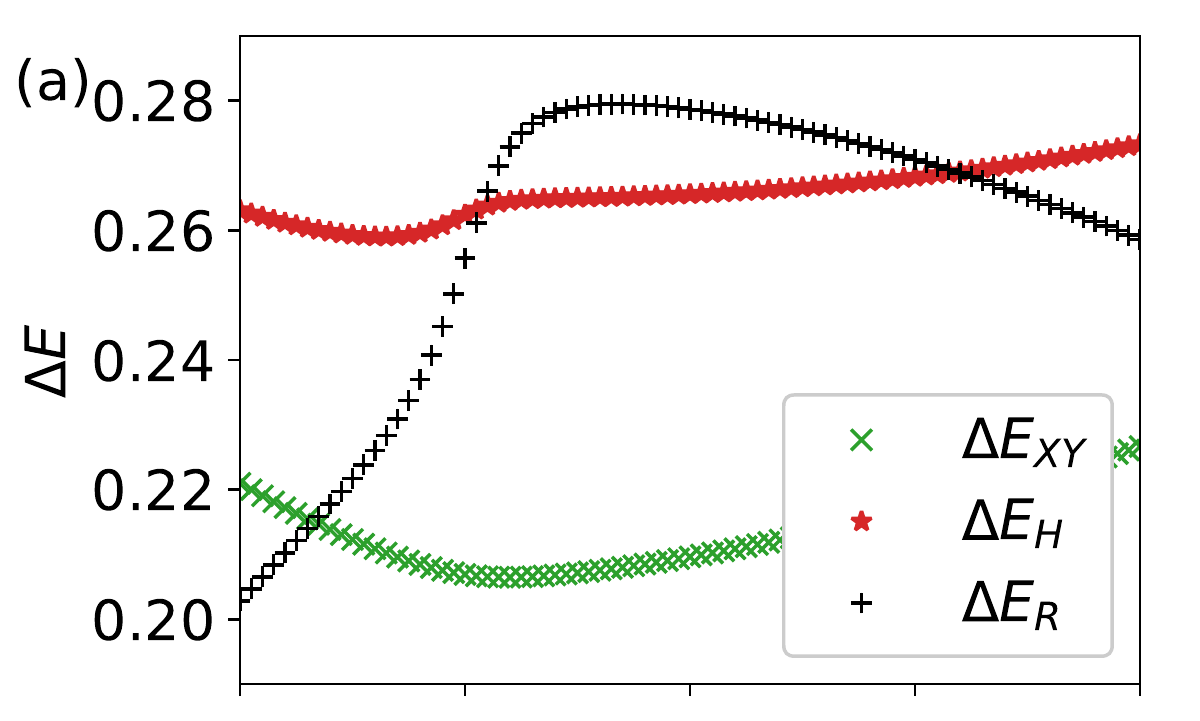}
\includegraphics[width=0.45\textwidth]{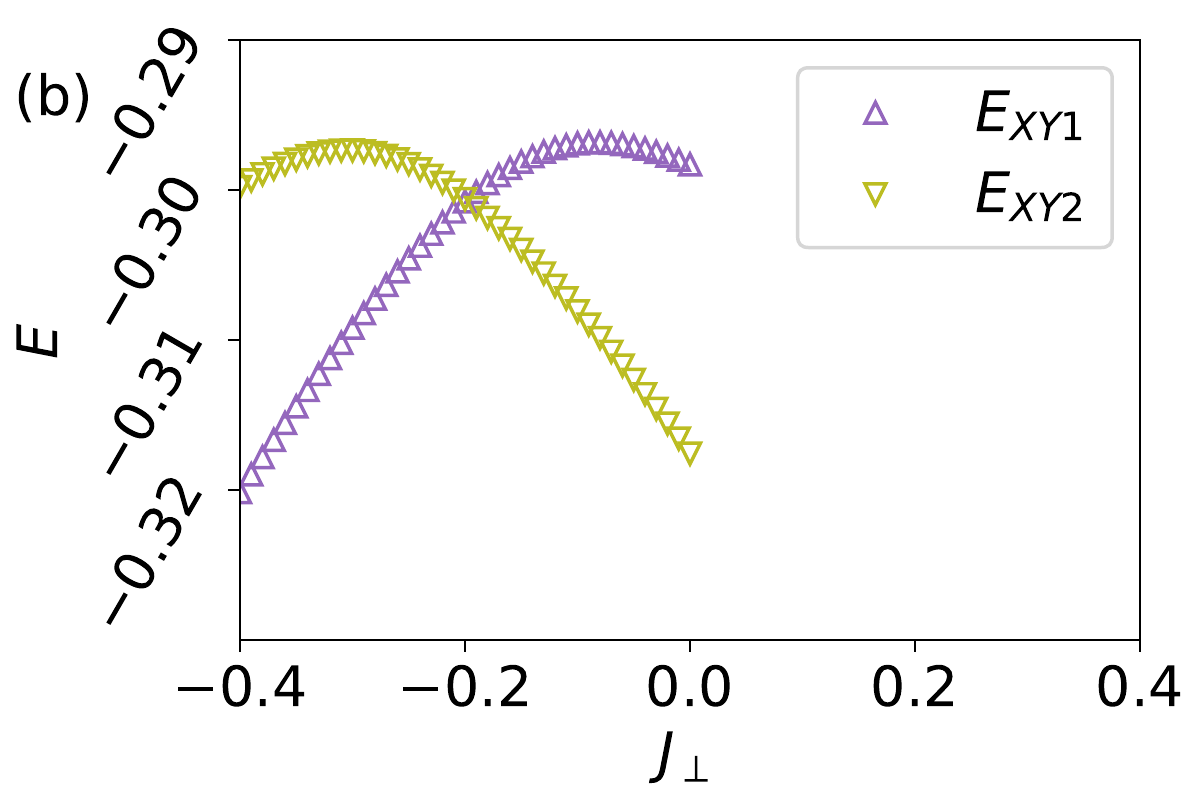}
\caption{
(a) Excitation energies $\Delta \Erm{XY}$, $\Delta\Erm{H}$, and $\Delta\Erm{R}$ in Eq.\ \eqref{eq:E_XYHR} with $L=12$ 
and (b) Eigenenergies $\Erm{XY1}$ and $\Erm{XY2}$ in Eq.\ \eqref{eq:E_XY1XY2} with $L=11$, 
calculated by exact diagonalization across the RS*-XY1-XY2 transition
for $\Delta=-0.2$ and $\Jdiag=-0.1$; see Fig.\ \ref{fig:phasediagram_Jxpm0.1}(b). 
The crossing of $\Delta\Erm{R}$ and $\Delta\Erm{XY}$ at $J_{\perp}=-0.335$ in (a) gives a finite-size estimate of the RS*-XY1 transition. 
The crossing of $\Erm{XY1}$ and $\Erm{XY2}$ at $J_\perp=-0.199$ in (b) gives a finite-size estimate of the XY1-XY2 transition. 
The crossing of $\Delta\Erm{H}$ and $\Delta\Erm{R}$ at $J_{\perp}=-0.185$ and $J_\perp=0.237$ indicate sign changes of $g_+$ or $y_2$ 
but do not indicate phase transitions as the $g_+$ or $y_2$ terms are irrelevant there. 
}
\label{fig:levelspec_Dm020}
\end{figure}

The key idea of the level spectroscopy is to relate the low-lying energy levels 
to the (running) coupling constants in the effective Hamiltonians in Eqs.\ \eqref{eq:Heff} and \eqref{eq:Hp_eff}. 
This method is well established for the sine-Gordon model in Eq.\ \eqref{eq:Hp_eff} 
\cite{Nomura_1995, Kitazawa_1997, Nomura_1998, PhysRevB.59.11358, cond-mat/0201072, PhysRevB.72.014449, PhysRevB.104.165132}; 
see, in particular, its application to a spin-$\frac12$ ladder by Hijii {\it et al.}\ \cite{PhysRevB.72.014449}. 
We consider a finite ladder of length $L$, 
and impose either the periodic boundary condition (PBC) ($\Sv_{n,L+j}=\Sv_{n,j}$) 
or the twisted boundary condition (TBC) [$S_{n,L+j}^{x,y}=-S_{n,j}^{x,y}$ and $S_{n,L+j}^z=S_{n,j}^z~(n=1,2; j=1,2,\dots)$]. 
The eigenstates can be classified by the total magnetization $S^z:=\sum_{n,j} S_{n,j}^z$, 
the wave vector $\qv=(q_x,q_y)$, and the bond-centered parity $P=\pm 1$. 
In the easy-plane regime where we apply the level spectroscopy, 
the ground state for even $L$ is in the sector with $S^z=0$, $\qv=\bm{0}$, and $P=1$. 
To investigate the Gaussian and BKT transitions described by the sine-Gordon model in Eq.\ \eqref{eq:Hp_eff}, 
we examine the following low-lying energy levels: 
\begin{subequations}\label{eq:E_XYHR}
\begin{align}
 \Erm{XY}&:~L=\even,~\PBC,~S^z=\pm 2,~\qv=\bm{0},~P=1 \label{eq:E_XY},\\
 \Erm{H}&:~L=\even,~\TBC,~S^z=0,~q_y=0,~P=-1, \label{eq:E_H}\\
 \Erm{R}&:~L=\even,~\TBC,~S^z=0,~q_y=0,~P=1. \label{eq:E_R}
\end{align}
\end{subequations}
For the TBC, $q_x$ is omitted as the system breaks the translational symmetry 
along the legs \footnote{
The system is still invariant under the translation followed by a gauge transformation. 
One may therefore introduce the wave vector for this symmetry. 
However, the introduction of such a wave vector is not important for the present analysis. 
}.
We subtract the ground-state energy for the PBC from these energies 
to obtain the excitation energies $\Delta E$. 
In the sine-Gordon model in Eq.\ \eqref{eq:Hp_eff}, 
the excitation energies are related to the running coupling constants $y_1(\ell)$ and $y_2(\ell)$ 
as \footnote{Because of differences in the bosonization conventions, the sign of $y_2$ is opposite between the present paper and Ref.\ \cite{PhysRevB.72.014449}.}
\begin{subequations}\label{eq:E_XYHR_y1y2}
\begin{align}
 \Delta \Erm{XY} &=\frac{2\pi v_+ }{La} \qty( \frac12-\frac{y_1(\ell)}{4} ),\\
 \Delta \Erm{H} &=\frac{2\pi v_+ }{La} \qty( \frac12+\frac{y_1(\ell)}{4}+\frac{y_2(\ell)}{2} ),\\
 \Delta \Erm{R} &=\frac{2\pi v_+ }{La} \qty( \frac12+\frac{y_1(\ell)}{4}-\frac{y_2(\ell)}{2} ),
\end{align}
\end{subequations}
where the logarithmic RG scale $\ell$ is related to the system size $L$ as $e^\ell=L/(2\pi)$. 
We can see from these relations that the crossing of $\Delta\Erm{H}$ and $\Delta\Erm{R}$ occurs at $y_2(\ell)=0$, 
which corresponds to the Haldane-RS* or Haldane*-RS Gaussian transition. 
Similarly, the crossing of $\Delta\Erm{XY}$ and $\Delta\Erm{R}$ occurs at $y_1(\ell)=y_2(\ell)$, 
which corresponds to the XY1-RS* or XY2-RS BKT transition. 
The crossing of $\Delta\Erm{XY}$ and $\Delta\Erm{H}$ occurs at $y_1(\ell)=-y_2(\ell)$, 
which corresponds to the XY1-Haldane or XY2-Haldane* BKT transition. 

The above argument is based on the sine-Gordon model \eqref{eq:Hp_eff} for the symmetric channel, 
and is therefore valid only when the $\theta_-$ field is locked tightly into the minimum of the $\gt_-$ term in Eq.\ \eqref{eq:Heff}. 
In the highly frustrated regime with $\Jperp\approx 2\Jdiag$, however, 
the bare value of $\gt_-$ is small in magnitude, and shows a sign change when varying $\Jperp-2\Jdiag$, as seen in Eq.\ \eqref{eq:coeff_gtm}; 
the reduction to the effective sine-Gordon model \eqref{eq:Hp_eff} may no longer be legitimate. 
In this regime, it is useful to relate the low-lying energy levels 
to the coupling constants in the two-channel effective field theory in Eq.\ \eqref{eq:Heff}; 
see Appendix \ref{app:spectra_eff} for detailed calculations. 
Specifically, we additionally examine the following low-lying energy levels: 
\begin{subequations}\label{eq:E_XY1XY2}
\begin{align}
 \Erm{XY1}&:~L=\odd,~\TBC,~S^z=0,~q_y=0,~P=1,\\
 \Erm{XY2}&:~L=\odd,~\TBC,~S^z=0,~q_y=\pi,~P=1.
\end{align}
\end{subequations}
The energy difference $\Erm{XY1}-\Erm{XY2}$ is related to the coupling constant $\gt_-$; see Eq.\ \eqref{eq:E_Mm1_cs}. 
Therefore, the XY1-XY2 Gaussian transition with $\gt_-=0$ can be determined from the crossing of $\Erm{XY1}$ and $\Erm{XY2}$. 
Similarly, $\Erm{H}-\Erm{R}$ is related to the coupling constant $g_+$; see Eq.\ \eqref{eq:E_Wp1_cs}. 

In Fig.\ \ref{fig:levelspec_D020}(a), we examine the excitation energies $\Delta\Erm{H}$ and $\Delta\Erm{R}$ 
as a function of $\Jperp$ for fixed $\Delta=0.2$ and $\Jdiag=-0.1$. 
We note that $\Delta\Erm{XY}$ (not shown) is always above these energies in the plotted parameter range. 
For $\Jperp$ away from $2\Jdiag=-0.2$, the relations in Eq.\ \eqref{eq:E_XYHR_y1y2} 
based on the sine-Gordon model \eqref{eq:Hp_eff} is expected to be appropriate. 
Therefore, the crossing of $\Delta\Erm{H}$ and $\Delta\Erm{R}$ at $J_{\perp,c}^\text{H*-R}=0.056$ 
gives a finite-size estimate of the Haldane*-RS transition point. 
We extrapolate the data of $J_{\perp,c}^\text{H*-R}(L)$ with a linear function of $1/L^2$,  
and obtain the estimate $J_{\perp,c}^\text{H*-R}(L\to\infty)=0.066$ in the thermodynamic limit, as shown in Fig.\ \ref{fig:levelspec_D020}(b) 
\footnote{While some deviation of the data from the line is seen in Fig.\ \ref{fig:levelspec_D020}(b), 
such deviation is rather specific to the regime $\Delta>0$ in Fig.\ \ref{fig:phasediagram_Jxpm0.1}(b), 
where finite-size effects occur in a nontrivial manner. 
In most of the other cases, we find that a linear function of $1/L^2$ can fit the data of the crossing points well; 
see, e.g., Fig.\ \ref{fig:levelspec_Jrm050_Jcm050} in Appendix \ref{app:num_data}.}. 
We note that similar extrapolation has been performed for BKT transitions in Refs.\ \cite{Nomura_1998,PhysRevB.72.014449}. 
Meanwhile, for $\Jperp\approx 2\Jdiag=-0.2$, it is more appropriate to relate $\Delta\Erm{H}-\Delta\Erm{R}$ 
to the coupling constant $g_+$ in the two-channel theory \eqref{eq:Heff}. 
Therefore, the crossing of $\Delta\Erm{H}$ and $\Delta\Erm{R}$ at $J_{\perp,c}=-0.179$ indicates a sign change of $g_+$. 
This sign change does not necessarily indicate a phase transition 
as various terms compete in the two-channel theory \eqref{eq:Heff}.
We may still use the crossing point of $\Delta\Erm{H}$ and $\Delta\Erm{R}$ 
as a rough estimate of the left boundary of the Haldane* phase because $g_+<0$ favors the formation of the Haldane* phase. 
A precise determination of this phase boundary is beyond the scope of this work. 

In Fig.\ \ref{fig:levelspec_Dm020}, we examine the energy levels in Eqs.\ \eqref{eq:E_XYHR} and \eqref{eq:E_XY1XY2} 
as a function of $\Jperp$ for fixed $\Delta=-0.2$ and $\Jdiag=-0.1$. 
The relations in Eq.\ \eqref{eq:E_XYHR_y1y2} can again be used for $\Jperp$ away from $2\Jdiag=-0.2$. 
Therefore, the crossing of $\Delta\Erm{R}$ and $\Delta\Erm{XY}$ at $J_{\perp}=-0.335$ in Fig.\ \ref{fig:levelspec_Dm020}(a) gives a finite-size estimate of the RS*-XY1 transition. 
Meanwhile, for $\Jperp\approx 2\Jdiag=-0.2$, we can analyze the data in terms of the two-channel theory \eqref{eq:Heff}. 
As the $\gt_-$ term has a much lower scaling dimension $(2K_-)^{-1}<(2K)^{-1}<1/2$ than other terms in this regime 
\footnote{
In this regime, the $g_-$, $\gamma_1$, and $\gamma_\bs$ terms with the scaling dimensions $2K_-$, $2K_++(2K_-)^{-1}$, and $2K_++2K_-$ 
are expected to be irrelevant. 
Furthermore, the $g_-$ term only has a small bare value for $\Jperp\approx 2\Jdiag$, as seen in Eq.\ \eqref{eq:coeff_cos_gpm}. 
}, 
we can expect that the long-distance physics in the antisymmetric channel is dominantly determined by this term. 
Therefore, the crossing of $\Erm{XY1}$ and $\Erm{XY2}$ at $J_\perp=-0.199$ in Fig.\ \ref{fig:levelspec_Dm020}(b) gives a finite-size estimate of the XY1-XY2 transition. 
The crossing of $\Delta\Erm{H}$ and $\Delta\Erm{R}$ at $J_{\perp}=-0.185$ and $0.237$ indicate sign changes of $g_+$ or $y_2$ 
but do not indicate phase transitions as the $g_+$ or $y_2$ terms are irrelevant there. 

In the phase diagrams for $\Jdiag=0$ and $\pm 0.1$ in Figs.\ \ref{fig:phasediagram_j_jp_model} and \ref{fig:phasediagram_Jxpm0.1}, 
the transition points in the easy-plane regime are determined by the level spectroscopy analysis as described above. 
We can confirm an overall consistency with the schematic phase diagrams in Fig.\ \ref{fig:phases_Delta0} obtained by the Abelian bosonization. 
However, we leave the issue of the detailed phase structure for $\Jperp\approx 2\Jdiag<0$ and $0<\Delta<1$ for a future work. 
In this regime, the $g_+$, $\gt_-$ and $\gamma_1$ terms in the effective Hamiltonian \eqref{eq:Heff} compete 
in a nontrivial manner as discussed in Sec.\ \ref{sec:bos_A_easyplane}; 
correspondingly, it is nontrivial to relate phase transitions to certain level crossings. 

Lastly, we compare our phase diagram of the simple XXZ ladder in Fig.\ \ref{fig:phasediagram_j_jp_model} 
with the previous tensor network study by Li {\it et al.} \cite{Li_2017}. 
In Ref.\ \cite{Li_2017}, the estimated ranges of the XY1 and XY2 phases are much broader than in Fig.\ \ref{fig:phasediagram_j_jp_model}. 
The analyses of Ref.\ \cite{Li_2017} are based on the calculations of the fidelity and (local and string) order parameters. 
In the theory of BKT transitions, it is known that the correlation length becomes extremely large and shows an essential singularity 
when the system approaches the transition point from the gapped phase \cite{Kosterlitz_1974}. 
Therefore, changes in physical quantities across a BKT transition are often obscured in numerical simulations. 
The level spectroscopy has been developed to overcome this difficulty by relating the transitions to certain level crossings 
using the knowledge of the underlying field theory and the RG flow 
(see, e.g., Refs.\ \cite{PhysRevB.86.094417, PhysRevB.91.014418} for other approaches in this direction). 
Based on the consistency with the bosonization analysis as well as a rather small dependence of the level crossing points on the system size, 
we conclude that Fig.\ \ref{fig:phasediagram_j_jp_model} gives an accurate phase diagram of the simple XXZ ladder 
which is improved from Ref.\ \cite{Li_2017}. 

\section{Summary and outlook}\label{sec:summary}

In this paper, we have studied the ground-state phase diagram of a spin-$\frac12$ frustrated XXZ ladder model \eqref{eq:ladfrust_XXZ}
by means of effective field theory, the iDMRG, and exact diagonalization.
An interesting feature of this model is the emergence of an inter-leg effective dimer attraction in the RG process;
in the isotropic model, this interaction induces the CD phase in the highly frustrated regime $\Jperp\approx 2\Jdiag$,
especially for ferromagnetic $\Jpd<0$ \cite{PhysRevLett.93.127202, PhysRevB.81.064432}.
By combining the bosonization analyses of Refs.\ \cite{PhysRevLett.93.127202, PhysRevB.81.064432, PhysRevB.82.214420, ogino2021spt},
we have argued that an interplay between the RG-generated dimer attraction and the XXZ anisotropy
brings about a rich phase structure as shown in Fig.\ \ref{fig:phases_Delta1}. 
Specifically, the Haldane-CD transition point in the isotropic model has turned out to be a crossing point of Gaussian and Ising transition lines in the XXZ model, 
and the SN and RS* phases appear between these lines. 
This phase structure is confirmed in the numerically obtained phase diagram for $\Jdiag/J=-0.5$ in Fig.\ \ref{fig:phasediagram}.
Furthermore, we have found that in the easy-plane regime $-1<\Delta<1$,
four gapped featureless phases and two critical phases 
compete in a complex manner depending on the signs of $\Jpd$, 
as shown in Figs.\ \ref{fig:phases_Delta0}(b,c) and \ref{fig:phasediagram_Jxpm0.1}.
This complex phase structure results from a competition among $g_+$, $\gt_-$, and $\gamma_1$ terms in the effective Hamiltonian \eqref{eq:Heff}.
We have also obtained an accurate phase diagram of the simple XXZ ladder model with $\Jdiag=0$, as shown in Fig.\ \ref{fig:phasediagram_j_jp_model}. 

Nontrivial effects of RG-generated effective interactions have been explored 
in a variety of frustrated Heisenberg magnets \cite{PhysRevLett.93.127202, PhysRevB.81.064432, 
PhysRevB.94.035154, PhysRevLett.98.077205, PhysRevB.84.174415, 2205.15525, 
PhysRevB.72.094416,PhysRevB.78.174420,PhysRevB.85.104417}. 
The coupling constants of such generated interactions are generally small in magnitude
if they are measured in the form of renormalized initial values as in Eq.\ \eqref{eq:gNE_initial}. 
However, highly frustrated magnets are often susceptible to weak perturbations, 
and the effects of the generated interactions can overcome other interactions existing at the bare level. 
This is an intriguing mechanism that can give rise to nontrivial phases which are difficult to conceive from the microscopic Hamiltonians. 
The present study demonstrates that this mechanism can produce an even more intricate phase structure in the presence of exchange anisotropy. 
In view of recent advances in numerical methods for efficiently dealing 
with two-dimensional frustrated quantum systems \cite{PhysRevX.11.031034,PhysRevB.94.035133,PhysRevB.94.155123,PhysRevX.9.031041, doi:10.7566/JPSJ.91.062001}, 
it will be interesting to test the roles of RG-generated interactions in two-dimensional systems 
(such as those in Refs.\ \cite{PhysRevLett.93.127202,PhysRevB.72.094416,PhysRevB.78.174420,PhysRevB.85.104417}) by state-of-the-art algorithms 
and to further explore a richer phase structure in the presence of exchange anisotropy. 
Further researches in this direction will contribute to more systematic understanding of highly frustrated quantum magnets.


\begin{acknowledgments}
The authors thank Naoki Kawashima and Hyun-Yong Lee for stimulating discussions, 
and Toshiya Hikihara for sharing the data of Ref.\ \cite{PhysRevB.81.064432}.
This research was supported by JSPS KAKENHI Grants No.\ JP18K03446 and No.\ JP20K03780. 
The numerical computations were performed on computers at the Supercomputer Center, 
the Institute for Solid State Physics (ISSP), the University of Tokyo. 
\end{acknowledgments}

\appendix

\section{Finite-size spectra of the effective Hamiltonians}\label{app:spectra_eff}

\newcommand{\rt}{\tilde{r}}
\newcommand{\Qt}{\tilde{Q}}
\newcommand{\pit}{\tilde{\pi}}

In this appendix, we calculate finite-size spectra of the effective Hamiltonians in Eqs.\ \eqref{eq:Heff} and \eqref{eq:Hp_eff} 
on the basis of a perturbation theory from the Gaussian Hamiltonians (see Refs.\ \cite{CARDY1986186, PhysRevB.79.245414, PhysRevB.81.094430} for related approaches). 
Our calculations for the sine-Gordon model \eqref{eq:Hp_eff} reproduce some of the previous results 
in Refs.\ \cite{Nomura_1995, Kitazawa_1997, Nomura_1998, PhysRevB.59.11358, cond-mat/0201072, PhysRevB.72.014449, PhysRevB.104.165132}, 
which relate Gaussian and BKT transitions to crossing of certain energy levels. 
We also show that the sign changes of the coupling constants $g_+$ and $\gt_-$ in the two-channel theory \eqref{eq:Heff} 
correspond to certain level crossings. 
These results give the theoretical basis of our level spectroscopy analysis in Sec.\ \ref{sec:LevelSpec}.  

We first examine possible windings of the original fields $\phi_n$ and $\theta_n$ on each leg ($n=1,2$); 
such windings lead to low-energy excitations as we will see later. 
As these fields are compactified on circles, they are allowed to change across the legs as 
\begin{subequations}\label{eq:phi_theta_n_winding}
\begin{align}
 \phi_n(x+La)&=\phi_n(x)+\sqrt{\pi} M_n, \label{eq:phi_n_winding}\\
 \theta_n(x+La)&=\theta_n(x)+2\sqrt{\pi} W_n. 
\end{align}
\end{subequations}
Here, $M_n$ is equal to the magnetization $\sum_j S_{n,j}^z$ on the $n$th leg, 
and $W_n$ is the winding number of the phase $\sqrt{\pi}\theta_n$. 
Possible values of $M_n$ and $W_n$ depend on the system size $L$ and the boundary condition. 
For even $L$ and the PBC, both $M_n$ and $W_n$ are integers. 
For even $L$ and the TBC, $M_n$ are integers while $W_n$ are half-integers. 
For odd $L$ and the TBC, $M_n$ are half-integers while $W_n$ are integers. 
Here, possible values of $W_n$ are determined from the following consideration: 
the phase of the term $e^{i\sqrt{\pi}\theta_n}b_0(-1)^j$ in Eq.\ \eqref{eq:Sp_bos} changes by $2\pi W_n+\pi L$ across the leg, 
and this change must be an even (odd) multiple of $\pi$ for the PBC (TBC). 

We now consider the Gaussian part of the effective Hamiltonian \eqref{eq:Heff}, i.e., the two-channel TLL. 
In this theory, the fields $\phi_\nu$ and $\theta_\nu$ ($\nu=\pm$) have the mode expansions
\begin{subequations}\label{eq:phi_theta_expand}
\begin{align}
 &\phi_\nu(x)=\phi_{\nu,0}+\Qt_\nu \frac{x}{La} \notag\\
 &+ \sum_{m\ne 0} \sqrt{\frac{K_\nu}{4\pi |m|}} e^{-{\alpha|k_m|}/{2}} \left( a_{m,\nu}e^{ik_m x} + a_{m,\nu}^\dagger e^{-ik_m x} \right),\\
 &\theta_\nu(x)=\theta_{\nu,0}+Q_\nu \frac{x}{La} \notag\\
 &- \sum_{m\ne 0} \frac{\mathrm{sgn}(m)e^{-{\alpha|k_m|}/{2}}}{\sqrt{4\pi K_\nu |m|}}  \left( a_{m,\nu}e^{ik_mx} + a_{m,\nu}^\dagger e^{-ik_mx} \right),
 \label{eq:theta_expand}
\end{align}
\end{subequations}
where $k_m:=2\pi m/La$ and $\alpha$ is a short-distance cutoff. 
The constant and winding parts in the expansions satisfy the commutation relations 
\begin{align}
 [\phi_{\nu,0},\theta_{\nu,0}]=-\frac{i}{2},~
 [\phi_{\nu,0},Q_{\nu'}]=[\theta_{\nu,0},\Qt_{\nu'}]=i\delta_{\nu\nu'}. 
\end{align}
To be consistent with Eq.\ \eqref{eq:phi_theta_n_winding}, $\Qt_\nu$ and $Q_\nu$ have the eigenvalues
$\sqrt{\pi/2} M_\nu$ and $\sqrt{2\pi} W_\nu$, respectively, where
\begin{align}
 M_\pm:=M_1\pm M_2,~~
 W_\pm:=W_1\pm W_2. 
\end{align}
We note that $M_\pm$ (likewise, $W_\pm$) are not independent of each other for a given system. 
For even $L$ and the PBC, for example, $M_\pm$ are both even or both odd since $M_++M_-=2M_1\in 2\mathbb{Z}$. 

By substituting Eq.\ \eqref{eq:phi_theta_expand} into the Gaussian part of Eq.\ \eqref{eq:Heff}, we obtain
\begin{align}
 H_\mathrm{Gauss}&=\sum_{\nu=\pm} \frac{v_\nu k_1}{4\pi} \qty( \frac{1}{K_\nu} \Qt_\nu^2 + K_\nu Q_\nu^2 ) \notag\\
 &+ \sum_{\nu=\pm} \sum_{m\ne 0} v_\nu |k_m|\qty( a_{m,\nu}^\dagger a_{m,\nu}+\frac12 ). 
\end{align}
The spectrum related to the zero modes are thus given by
\begin{align}\label{eq:E_Mpm_Wpm}
 \Delta E(M_\pm,W_\pm)=\sum_{\nu=\pm} v_\nu k_1 \qty( \frac{1}{8K_\nu} M_\nu^2+\frac{K_\nu}{2} W_\nu^2 ).
\end{align}
For even $L$ and the PBC, the corresponding eigenstates are given by
\begin{align}\label{eq:ket_Mpm_Wpm}
 \ket{M_\pm,W_\pm}= \exp[i\sum_{\nu=\pm} \qty(\sqrt{\frac{\pi}{2}} M_\nu \theta_{\nu,0} + \sqrt{2\pi} W_\nu\phi_{\nu,0}) ] \ket{0},
\end{align}
where $\ket{0}$ is the ground state satisfying $M_\nu=W_\nu=0$ and $a_{m,\nu}\ket{0}=0~(\nu=\pm;m\ne 0)$. 
We may use Eq.\ \eqref{eq:ket_Mpm_Wpm} for the other cases (odd $L$ or the TBC) as well;  
however, $\ket{0}$ should then be understood as the fictitious state with unphysical winding numbers.  
In the spectrum in Eq.\ \eqref{eq:E_Mpm_Wpm}, we are particularly interested in the following energy levels: 
\begin{subequations}
\begin{align}
 \Delta E(M_+=\pm 2)&=\frac{v_+k_1}{2K_+}~~(L=\text{even,~PBC}), \label{eq:E_Mp2}\\
 \Delta E(W_+=\pm 1) &=\frac{v_+k_1K_+}{2}~~(L=\text{even,~TBC}),\label{eq:E_Wp1}\\
 \Delta E(M_-=\pm 1) &= \frac{v_-k_1}{8K_-}~~(L=\text{odd,~TBC}),\label{eq:E_Mm1}
\end{align}
\end{subequations}
where only nonzero winding numbers are shown in the argument of $\Delta E(\cdot)$. 
Each of these levels is doubly degenerate in the Gaussian Hamiltonian. 

We now consider the effects of perturbations in the effective Hamiltonian \eqref{eq:Heff}. 
The $\gt_-$ term has a nonzero matrix element between the states $\ket{M_-=\pm 1}$ as we show in the following. 
Using the expansion \eqref{eq:theta_expand}, this matrix element can be expressed as
\begin{equation}\label{eq:matel_gtm}
\begin{split}
 &\bra{M_-=-1} \gt_- \cos(\sqrt{2\pi}\theta_-) \ket{M_-=1} \\
 &=\frac{\gt_-}2 \bra{0} e^{i\sqrt{\frac{\pi}{2}}\theta_{-,0}} e^{-i\sqrt{2\pi}\theta_-(x)} e^{i\sqrt{\frac{\pi}{2}} \theta_{-,0}} \ket{0}\\
 &=\frac{\gt_-}2 \bra{0} e^{i[A(x)+A^\dagger(x)]} \ket{0},
\end{split}
\end{equation}
where we introduce the short-hand notation 
\begin{equation}
 A(x)=\sum_{m\ne 0} \frac{\mathrm{sgn}(m)  e^{-{\alpha|k_m|}/{2}} }{\sqrt{2K_- |m|}} a_{m,-}e^{ik_mx}. 
\end{equation}
Equation \eqref{eq:matel_gtm} can further be calculated as \cite{PhysRevB.81.094430}
\begin{equation}\label{eq:matel_gtm_calc}
 \frac{\gt_-}2 e^{-[A,A^\dagger]/2} \bra{0} e^{iA^\dagger} e^{iA} \ket{0}\\
 \approx \frac{\gt_-}2 (k_1\alpha)^{1/(2K_-)}, 
\end{equation}
where we use $A(x)\ket{0}=0$ and
\begin{align}
 [A(x), A^\dagger(x)] &= \frac1{2K_-} \sum_{m\ne 0} \frac{e^{-\alpha |k_m|}}{|m|} 
 = - \frac1{K_-} \ln \qty( 1-e^{-k_1\alpha}) \notag\\
 &\approx  - \frac1{K_-}  \ln (k_1\alpha).
\end{align}
Therefore, the level \eqref{eq:E_Mm1} is split in the first-order perturbation theory. 
The new eigenstates are given by 
\begin{subequations}\label{eq:Mp1_cs}
\begin{align}
 \ket{|M_-|=1, \cos}&=\sqrt{2} \cos( \sqrt{\frac{\pi}{2}} \theta_{-,0} ) \ket{0}, \label{eq:Mp1_c}\\
 \ket{|M_-|=1, \sin}&=\sqrt{2} \sin( \sqrt{\frac{\pi}{2}} \theta_{-,0} ) \ket{0}, \label{eq:Mp1_s}
\end{align}
\end{subequations}
which have the eigenenergies 
\begin{align}\label{eq:E_Mm1_cs}
 \Delta E(|M_-|=1,\cos / \sin )
 = \frac{v_-k_1}{8K_-}\pm \frac{\gt_- La}{2}\qty(k_1\alpha)^{1/(2K_-)}. 
\end{align}
These correspond to the energy levels $\Erm{XY1}$ and $\Erm{XY2}$ in Eq.\ \eqref{eq:E_XY1XY2}. 
By examining the difference between these levels, we can probe the sign of $\gt_-$. 

Similarly, the $g_+$ term has the matrix element 
\begin{align}\label{eq:matel_gp}
  \bra{W_+=-1} g_+\cos(2\sqrt{2\pi}\phi_+) \ket{W_+=1}
 = \frac{g_+}2 \qty(k_1\alpha)^{2K_+}.
\end{align}
Therefore, the level \eqref{eq:E_Wp1} is split by this term. 
The new eigenstates are given by
\begin{subequations}\label{eq:Wp1_cs}
\begin{align}
 \ket{|W_+|=1,\cos}&= \sqrt{2} \cos( \sqrt{2\pi} \phi_{+,0} ) \ket{0}, \label{eq:Wp1_c}\\
 \ket{|W_+|=1,\sin}&= \sqrt{2} \sin( \sqrt{2\pi} \phi_{+,0} ) \ket{0}, \label{eq:Wp1_s}
\end{align}
\end{subequations}
which have the eigenenergies 
\begin{equation}\label{eq:E_Wp1_cs}
 \Delta E(|W_+|=1,\cos / \sin ) 
 =\frac{v_+k_1 K_+}{2}\pm \frac{g_+ La}{2}\qty(k_1\alpha)^{2K_+}. 
\end{equation}
These levels correspond to $\Erm{H}$ and $\Erm{R}$ in Eq.\ \eqref{eq:E_XYHR}. 
The difference between these levels is thus related to the coupling constant $g_+$. 
Meanwhile, the level \eqref{eq:E_Mp2} remains unchanged in the first-order perturbation theory, 
and corresponds to $\Erm{XY}$ in Eq.\ \eqref{eq:E_XY}. 

We next consider the effective sine-Gordon Hamiltonian \eqref{eq:Hp_eff}, 
which is obtained from Eq.\ \eqref{eq:Heff} when $\theta_-$ is locked into the minimum of the $\gt_-$ term. 
Equations \eqref{eq:E_Mp2} and \eqref{eq:E_Wp1_cs} are still valid in this case 
if $g_+$ is replaced by $g_+'$. 
We consider the regime with $K_+= 1+y_1/2~(|y_1|\ll 1)$, where the BKT transition is expected to occur. 
In this case, Eq.\ \eqref{eq:E_Mp2} can be approximated as
\begin{align}\label{eq:E_Mp2_y1}
 \Delta E(M_+=\pm 2)=v_+k_1 \qty( \frac12 - \frac{y_1}{4} ). 
\end{align}
Meanwhile, Eq.\ \eqref{eq:E_Wp1_cs} can be written as
\begin{align}\label{eq:E_Wp1_cs_yy}
 \Delta E(|W_+|=1,\cos / \sin ) 
 = v_+k_1 \qty[ \frac12 + \frac{y_1}{4} \pm \frac{y_2}{2} (k_1\alpha)^{y_1} ]. 
\end{align}
We have so far used the bare coupling constants $y_1$ and $y_2$. 
For $|y_2|\ll y_1$, the running coupling constants behave as 
\begin{equation}
 y_1(\ell)=y_1,~~y_2(\ell)=y_2 e^{-y_1\ell}. 
\end{equation}
Therefore, Eqs.\ \eqref{eq:E_Mp2_y1} and \eqref{eq:E_Wp1_cs_yy} can be rewritten as Eq.\ \eqref{eq:E_XYHR_y1y2} 
under the correspondence $e^\ell=(k_1\alpha)^{-1}=La/(2\pi \alpha)$. 
Based on a scaling analysis, one can further argue that Eq.\ \eqref{eq:E_XYHR_y1y2} holds 
near the multicritical point $(y_1,y_2)=(0,0)$, without the restriction $|y_2|\ll y_1$ \cite{PhysRevB.81.094430}. 

Let us determine the quantum numbers of the states discussed above. 
Below we assume that the (fictitious) state $\ket{0}$ is in the sector with $S^z=0$, $\qv=\bm{0}$, and $P=1$. 
Based on the spin-field correspondence in Eq.\ \eqref{eq:Spin_dim_bos}, the states $\ket{M_+=\pm 2}$ can be rewritten approximately as 
\begin{align}
 \ket{M_+=\pm 2}=e^{\pm \sqrt{2\pi}\theta_{+,0}} \ket{0}\sim \sum_j S_{1,j}^\pm S_{2,j}^\pm \ket{0}. 
\end{align}
It is clear from this expression that these states have the quantum numbers in Eq.\ \eqref{eq:E_XY}. 
Based again on Eq.\ \eqref{eq:Spin_dim_bos}, the reflection $\Sv_{n,j}\to \Sv_{n,L+1-j}$ can be described as 
\begin{subequations}
\begin{align}
 2\sqrt{\pi}\phi_n(x) &\to -2\sqrt{\pi}\phi_n(x')+\pi \chi_L,\\
 \sqrt{\pi}\theta_n(x) &\to \sqrt{\pi}\theta_n(x')+\pi \chi_L,
\end{align}
\end{subequations}
where $x'=La+a-x$ and $\chi_L:=[1+(-1)^L]/2$. From these, we have
\begin{subequations}
\begin{align}
 \sqrt{2\pi}\phi_+(x) &\to -\sqrt{2\pi}\phi_+(x')+\pi \chi_L,\\
 \sqrt{\frac{\pi}{2}}\theta_-(x) &\to \sqrt{\frac{\pi}{2}}\theta_-(x').
\end{align}
\end{subequations}
The uniform component $\sqrt{2\pi}\phi_{+,0}$ and $\sqrt{\pi/2}\theta_{-,0}$ transform in a similar manner. 
Therefore, the states in Eqs.\ \eqref{eq:Wp1_c} and \eqref{eq:Wp1_s} have $P=-1$ and $P=1$, respectively, 
while the states in Eq.\ \eqref{eq:Mp1_cs} have $P=1$. 
The interchange of the two legs can be described as $\phi_\pm(x)\to\pm\phi_\pm(x)$ and $\theta_\pm(x)\to\pm\theta_\pm(x)$. 
We can then determine the momentum $q_y$ in the $y$ direction in a similar manner; for example, $q_y=\pi$ for the state \eqref{eq:Mp1_s}. 
In these ways, we obtain all the quantum numbers in Eqs.\ \eqref{eq:E_XYHR} and \eqref{eq:E_XY1XY2}. 

In closing, we note that the quantum numbers of $\Erm{H}$ and $\Erm{R}$ in Eq.\ \eqref{eq:E_XYHR} 
can also be understood from the representative ground-state wave functions under the TBC. 
The rung-factorized RS and RS* states in Fig.\ \ref{fig:ladder_Jpd} are expected to give representative wave functions under the TBC as well; 
these states clearly have the quantum numbers in Eq.\ \eqref{eq:E_R}. 
The representative state in the Haldane phase is given by a valence bond solid state of effective spin-1's on the rungs; 
under the TBC, however, we have a twisted valence bond $(\ket{\uparrow\downarrow}+\ket{\downarrow\uparrow})/\sqrt{2}$ 
between the $1$st and $L$th rungs \cite{PhysRevB.59.11358}. 
Under the reflection, the $L-1$ valence bonds change the signs, while the twisted one does not; therefore we have $P=(-1)^{L-1}=-1$. 
It is also clear that the representative Haldane state has $q_y=0$ as the state on every rung is symmetrized. 
The Haldane* state also has $q_y=0$ and $P=-1$ 
as these quantum numbers remain unchanged under the action of $U_1^z$ on the zero-magnetization state. 
For $q_y$, this can be shown by using the identity $\Sigma_\sigma U_1^z \Sigma_\sigma=U_2^z=U_1^z \exp(i\pi S^z)$, 
where $U_n^z:={\exp} (i\pi \sum_j S_{n,j}^z)$ and $\Sigma_\sigma$ is the leg interchange. 

\section{Supplemental numerical data}\label{app:num_data}

\begin{figure}
\includegraphics[width=0.45\textwidth]{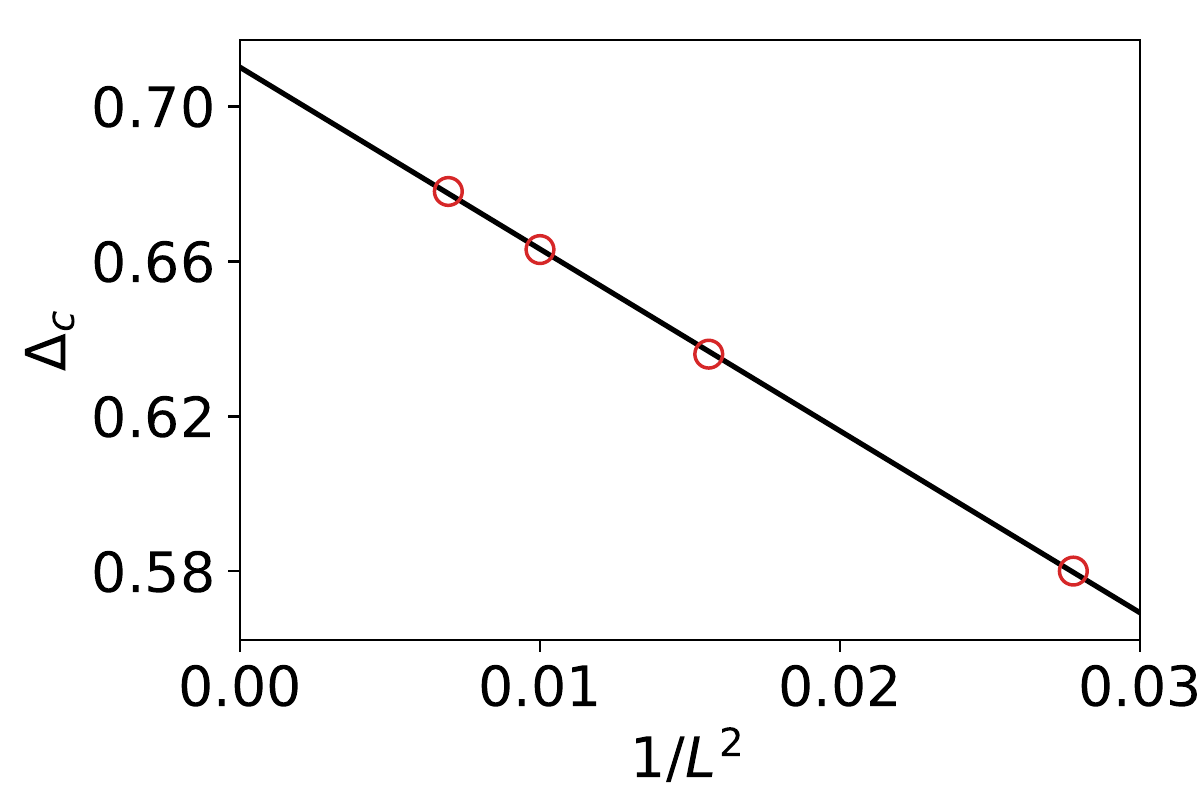}
\caption{\label{fig:levelspec_Jrm050_Jcm050}
Finite-size estimate $\Delta_c(L)$ of the Haldane*-RS Gaussian transition point 
obtained by the level spectroscopy for $\Jperp=\Jdiag=-0.5$; see Fig.\ \ref{fig:phasediagram}. 
Here, $\Delta_c(L)$ is obtained as the crossing point of $\Delta\Erm{H}$ and $\Delta\Erm{R}$; 
see Fig.\ \ref{fig:levelspec_D020}(a) for a similar analysis. 
The data of $\Delta_c(L)$ are fitted very well by a linear function of $1/L^2$. 
An extrapolation to the thermodynamic limit gives $\Delta_c(L\rightarrow\infty)=0.710$. 
}
\end{figure}

\begin{figure}
\includegraphics[width=90mm]{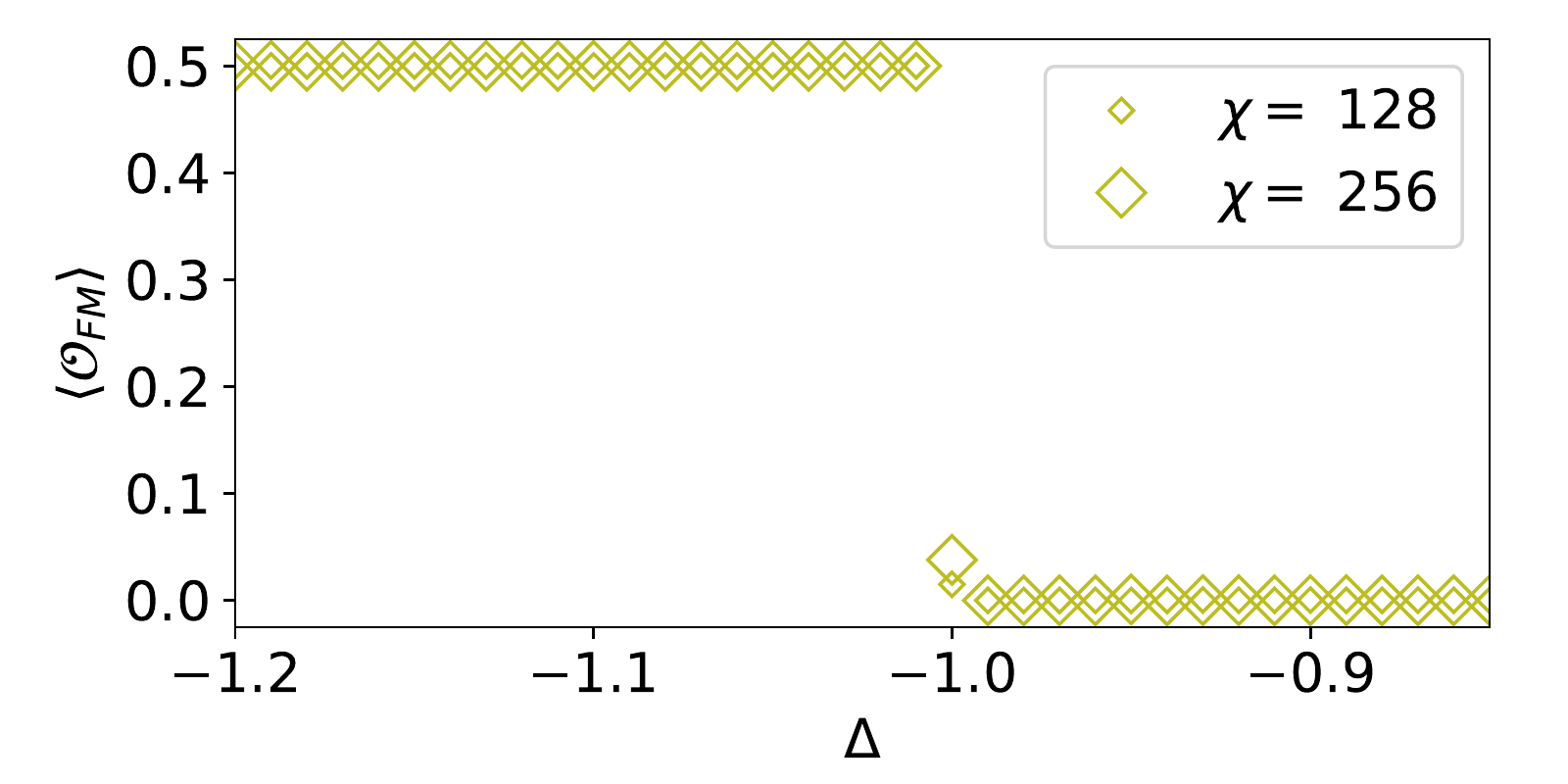}
\caption{\label{fig:order_FM_XY2}
FM order parameter \eqref{eq:O_FM}, i.e., the magnetization per site, 
calculated by the iDMRG across the FM-XY2 transition for $\Jperp=1$ and $\Jdiag=0$; see Fig.\ \ref{fig:phasediagram_j_jp_model}. 
The data are consistent with a first-order phase transition at the ferromagnetic SU(2) point $\Delta=-1$. 
}
\end{figure}

In this appendix, we provide some supplemental numerical data that support our analyses. 

To obtain the phase diagram for $\Jdiag=-0.5$ in Fig.\ \ref{fig:phasediagram}, which best illustrates our main results, 
we have mainly used the iDMRG method as explained in Sec.\ \ref{sec:iDMRG}. 
For the determination of the Haldane*-RG Gaussian transition points in this diagram, however, 
we have performed the level spectroscopy analysis to obtain accurate estimates. 
As an example, Fig.\ \ref{fig:levelspec_Jrm050_Jcm050} shows the finite-size estimate $\Delta_c(L)$ of the Haldane*-RS transition point 
obtained by the level spectroscopy for $\Jperp=\Jdiag=-0.5$. 
An extrapolation to the thermodynamic limit gives $\Delta_c(L\rightarrow\infty)=0.710$; 
this estimate of the transition point is consistent with the sudden changes 
in the topological indices between $\Delta=0.7$ and $0.8$ in Fig.\ \ref{fig:topo}(b). 
Figure \ref{fig:levelspec_Jrm050_Jcm050} also gives a successful example in which the data of crossing points 
are fitted very well by a linear function of $1/L^2$. 

In Sec.\ \ref{sec:Orders}, we have presented iDMRG results on the N\'eel, SN, and CD order parameters. 
In Fig.\ \ref{fig:order_FM_XY2}, we show an iDMRG result on the FM order parameter 
\begin{equation}\label{eq:O_FM}
 \expval{\mathcal{O}_{\text{FM}}(j)} = \frac14 \expval{S^z_{1,j} + S^z_{2,j} + S^z_{1,j+1} + S^z_{2,j+1}}
\end{equation}
across the FM-XY2 transition for $\Jperp=1$ and $\Jdiag=0$; see Fig.\ \ref{fig:phasediagram_j_jp_model}. 
Here, the iDMRG was performed without the constraint on the total magnetization $S^z=\sum_{n,j} S^z_{n,j}$. 
The data show a sudden change from the saturated value $1/2$ to $0$ across $\Delta=-1$, 
indicating a first-order phase transition. 
This is consistent with the following picture explained in Sec.\ \ref{sec:bos_A_Deltam1}: 
the levels with different magnetization $S^z$ cross at the ferromagnetic SU(2) point $\Delta=-1$ 
so that $S^z$ of the ground state changes abruptly. 


\bibliography{references}


\end{document}